\documentclass[apjl]{emulateapj}
\usepackage{psfig,amsfonts,amsmath,graphicx,natbib}
\def\nod{\nodata}

\def\pri{1}
\def\cfa{2}
\def\ein{3}

\shorttitle{Electromagnetic Counterparts of Neutron Star Mergers}
\shortauthors{Metzger \& Berger}

\begin{document}

\normalsize

\title{What is the Most Promising Electromagnetic Counterpart of a
Neutron Star Binary Merger?}

\author{
B.~D.~Metzger\altaffilmark{\pri,}\altaffilmark{\ein} 
and E.~Berger\altaffilmark{\cfa}
}

\altaffiltext{\pri}{Department of Astrophysical Sciences, Peyton Hall,
Princeton University, Princeton, NJ 08544}

\altaffiltext{\cfa}{Harvard-Smithsonian Center for Astrophysics, 60
Garden Street, Cambridge, MA 02138}

\altaffiltext{\ein}{NASA Einstein Fellow}

\begin{abstract} The inspiral and coalescence of double neutron star
(NS-NS) and neutron star-black hole (NS-BH) binaries are likely to be
detected by advanced networks of ground-based gravitational wave (GW)
interferometers.  Maximizing the science returns from such a discovery
will require the identification and localization of an electromagnetic
(EM) counterpart.  Here we critically evaluate and compare several
possible counterparts, including short-duration gamma-ray bursts
(SGRBs), ``orphan'' optical and radio afterglows, and $\sim\,$day-long
optical transients powered by the radioactive decay of heavy nuclei
synthesized in the merger ejecta (``kilonovae'').  We assess the
promise of each potential counterpart in terms of four ``Cardinal
Virtues'': detectability, high fraction, identifiability, and
positional accuracy.  For viewing angles within the half-opening angle
of the jet ($\theta_{\rm obs}\lesssim \theta_{j}$) the SGRB and
associated afterglow are easily detectable within the range of
Advanced LIGO/Virgo if the jet energy ($E_{j}$) and the circumburst
density ($n$) are similar to those inferred from existing SGRB
observations.  For modest off-axis angles ($\theta_{\rm obs}\lesssim
2\theta_{j}$), the orphan optical afterglow is detectable with LSST if
$E_{j,50}\,n_0^{7/8}\gtrsim 0.002$; the fraction of such events is
$\sim 7\theta_j^2\sim 0.1$.  At even larger viewing angles (i.e., the
majority of observers) the isotropic kilonova emission dominates, with
a peak optical brightness of $\sim 19-22$ mag within the Advanced
LIGO/Virgo volume, detectable with LSST using a specialized 1-day
cadence.  Radio afterglow emission from an initially off-axis jet or
from sub-relativistic ejecta is also isotropic, but peaks on a
timescale of months-years; this signal is detectable provided that
$E_{j,50}\,n_0^{7/8}\,(v/c)^{11/4}\gtrsim 0.2$ (for off-axis
afterglows, $v/c\sim 1$).  However, existing SGRB afterglows do not
satisfy this criterion, indicating a low probability of radio
detections.  Taking into account the search strategy for typical error
regions of tens of square degrees, our primary conclusion is that
SGRBs are the most useful EM counterparts to confirm the cosmic origin
of a few GW events, and to test the association with NS-NS/NS-BH
mergers.  However, for the more ambitious goal of localizing and
obtaining redshifts for a large sample of GW events, kilonovae are
instead preferred.  Off-axis optical afterglows will be detectable for
at most $\sim 10\%$ of all events, while radio afterglows are
promising only for the unique combination of energetic relativistic
ejecta in a high density medium, and even then will require hundreds
of hours of EVLA time per event spread over months-years.  Our main
recommendations from this analysis are: (i) an all-sky $\gamma$-ray
satellite is essential for temporal coincidence detections, and for GW
searches of $\gamma$-ray triggered events; (ii) LSST should adopt a
1-day cadence follow-up strategy, ideally with $\sim 0.5$ hr per
pointing to cover GW error regions (the standard 4-day cadence and
depth will severely limit the probability of a unique identification);
and (iii) radio searches should only focus on the relativistic case,
which requires observations for a few months. \end{abstract}
  

\section{Introduction} 
\label{sec:intro}
  
The first direct detection of gravitational waves (GWs) is anticipated
within the decade once the ground-based interferometers
LIGO\footnotemark\footnotetext{\url{http://www.ligo.caltech.edu}}
(\citealt{Abramovici+92}; \citealt{Abbott+09}) and
Virgo\footnotemark\footnotetext{\url{http://www.virgo.infn.it}}
(\citealt{Caron+99}; \citealt{Acernese+09}) are upgraded to
``advanced'' sensitivity (hereafter ALIGO/Virgo).  The Large Scale
Cryogenic Gravitational Wave Telescope (LCGT; \citealt{Kuroda+10}) is
under construction in Japan and is anticipated to join ALIGO/Virgo by
about 2018.  The most promising astrophysical GW sources in the
frequency range of these detectors are the inspiral and coalescence of
compact object binaries with neutron star (NS) and/or black hole (BH)
constituents.  Although this accomplishment will stand on its own
merits, optimizing the science returns from a GW detection will
require the identification and study of coincident electromagnetic
(EM) counterparts
(e.g.,~\citealt{Schutz86,Schutz02,Sylvestre03,Stubbs08,Phinney09,Stamatikos+09}).
This is important for several reasons, including lifting degeneracies
associated with the inferred binary parameters \citep{Hughes&Holz03};
reducing the signal-to-noise ratio for a confident GW detection
\citep{Kochanek&Piran93,Dalal+06,Harry&Fairhurst11}; and identifying
the merger redshift, thereby setting the energy scale and allowing an
independent measurement of the Hubble constant or other cosmological
parameters
\citep[e.g.][]{Krolak&Schutz87,Chernoff&Finn93,Holz&Hughes05,Deffayet&Menou07,Nissanke+10}.
The potential wealth of complementary information encoded in the EM
signal is likewise essential to fully unraveling the astrophysical
context of the event \citep{Phinney09,Mandel&Oshaughnessy10}, for example an association
with specific stellar populations (e.g., \citealt{Fong+10}).

Motivated by the importance of EM detections, in this paper we address
the critical question: {\it What is the most promising EM counterpart
of a compact object binary merger?}  The answer of course depends on
the definition of ``most promising''.  In our view, a promising
counterpart should exhibit four Cardinal Virtues, namely it should:

\begin{enumerate}
\item{Be detectable with present or upcoming telescope facilities,
provided a reasonable allocation of resources.}
\item{Accompany a high fraction of GW events.}
\item{Be unambiguously identifiable (a ``smoking gun''), such that it
can be distinguished from other astrophysical transients.}
\item{Allow for a determination of $\sim\,$arcsecond sky positions.}
\end{enumerate}

Virtue \#1 is necessary to ensure that effective EM searches indeed
take place for a substantial number of GW triggers.  Virtue \#2 is
important because a large number of events may be necessary to build
up statistical samples, particularly if GW detections are rare; in
this context, ALIGO/Virgo is predicted to detect NS-NS mergers at a
rate ranging from $\sim 0.4$ to $\sim 400$ yr$^{-1}$, with a
``best-bet'' rate of $\sim 40$ yr$^{-1}$ (\citealt{Abadie+10};
cf.~\citealt{Kopparapu+08}), while the best-bet rate for detection of
NS-BH mergers is $\sim 10$ yr$^{-1}$.  Virtue \#3 is necessary to make
the association with high confidence and hence to avoid contamination
from more common transient sources (e.g., supernovae).  Finally,
Virtue \#4 is essential to identifying the host galaxy and hence the
redshift, as well as other relevant properties (e.g., association with
specific stellar populations).

It is important to distinguish two general strategies for connecting
EM and GW events.  One approach is to search for a GW signal following
an EM trigger, either in real time or at a post-processing stage
(e.g., \citealt{Finn+99}; \citealt{Mohanty+04}).  This is particularly
promising for counterparts predicted to occur in temporal coincidence
with the GW chirp, such as short-duration gamma-ray bursts (SGRBs).
Unfortunately, most other promising counterparts (none of which have
yet been independently identified) occur hours to months after
coalescence\footnotemark\footnotetext{Predicted EM counterparts that
may instead {\it precede} the GW signal include emission powered by
the magnetosphere of the NS (e.g.~\citealt{Hansen&Lyutikov01};
\citealt{McWilliams&Levin11}), or cracking of the NS crust due to
tidal interactions (e.g.~\citealt{Troja+10}), during the final
inspiral.  However, given the current uncertainties in these models,
we do not discuss them further.}.  Thus, the predicted arrival time of
the GW signal will remain uncertain, in which case the additional
sensitivity gained from this information is significantly reduced.
For instance, if the time of merger is known only to within an
uncertainty of $\sim\,{\rm hours(weeks)}$, as we will show is the case
for optical(radio) counterparts, then the number of trial GW templates
that must be searched is larger by a factor $\sim 10^{4}-10^{6}$ than
if the merger time is known to within seconds, as in the case of
SGRBs.

\begin{figure}
\centerline{\psfig{file=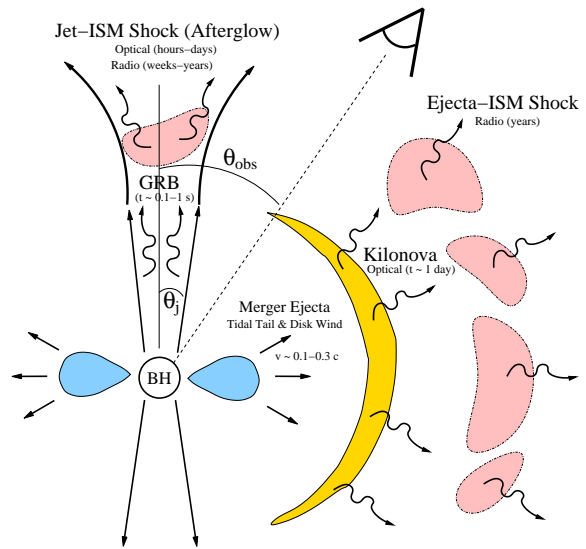,width=3in,angle=0}}
\caption[]{Summary of potential electromagnetic counterparts of
NS-NS/NS-BH mergers discussed in this paper, as a function of the
observer angle, $\theta_{\rm obs}$.  Following the merger a
centrifugally supported disk ({\it blue}) remains around the central
compact object (usually a BH).  Rapid accretion lasting $\lesssim 1$ s
powers a collimated relativistic jet, which produces a short-duration
gamma-ray burst (\S\ref{sec:GRB}).  Due to relativistic beaming, the
gamma-ray emission is restricted to observers with $\theta_{\rm
obs}\lesssim \theta_j$, the half-opening angle of the jet.
Non-thermal afterglow emission results from the interaction of the jet
with the surrounding circumburst medium ({\it red}).  Optical
afterglow emission is observable on timescales up to $\sim$
days$-$weeks by observers with viewing angles of $\theta_{\rm
obs}\lesssim 2\theta_j$ (\S\ref{sec:oa}).  Radio afterglow emission is
observable from all viewing angles (isotropic) once the jet
decelerates to mildly relativistic speeds on a timescale of
weeks-months, and can also be produced on timescales of years from
sub-relativistic ejecta (\S\ref{sec:ra}).  Short-lived isotropic
optical emission lasting $\sim\,$few days (kilonova; yellow) can also
accompany the merger, powered by the radioactive decay of heavy
elements synthesized in the ejecta (\S\ref{sec:kilonova}).}
\label{fig:cartoon}
\end{figure}

A second approach, which is the primary focus of this paper, is EM
follow-up of GW triggers.  A potential advantage in this case is that
counterpart searches are restricted to the nearby universe, as
determined by the ALIGO/Virgo sensitivity range (redshift $z\lesssim
0.05-0.1$).  On the other hand, a significant challenge are the large
error regions, which are estimated to be tens of square degrees even
for optimistic configurations of GW detectors (e.g.,
\citealt{Gursel&Tinto89}; \citealt{Fairhurst09}; \citealt{Wen&Chen10};
\citealt{Nissanke+11}).  Although it has been argued that this
difficulty may be alleviated if the search is restricted to galaxies
within 200 Mpc \citep{Nuttall&Sutton10}, we stress that the number of
galaxies with $L\gtrsim 0.1\,L^*$ (typical of SGRB host galaxies;
\citealt{Berger09,Berger11}) within an expected GW error region is $\sim
400$, large enough to negate this advantage for most search
strategies.  In principle the number of candidate galaxies could be
reduced if the distance can be constrained from the GW signal;
however, distance estimates for individual events are rather
uncertain, especially at that low SNRs that will characterize most
detections \citep{Nissanke+10}.  Moreover, current galaxy catalogs are
incomplete within the ALIGO/Virgo volume
(e.g.~\citealt{Kulkarni&Kasliwal09}), especially at lower
luminosities.  Finally, some mergers may also occur outside of their
host galaxies \citep{Berger10,Kelley+10}.

At the present there are no optical or radio facilities that can
provide all-sky coverage at a cadence and depth matched to the
expected light curves of EM counterparts.  As we show in this paper,
even the Large Synoptic Survey Telescope (LSST), with a planned
all-sky cadence of 4 d and a depth of $r\approx 24.7$ mag, is unlikely
to effectively capture the range of expected EM counterparts.  Thus,
targeted follow-up of GW error regions is required, whether the aim is
to detect optical or radio counterparts.  Even with this approach, the
follow-up observations will still require large field-of-view
telescopes to cover tens of square degrees; targeted observations of
galaxies are unlikely to substantially reduce the large amount of time
to scan the full error region.

Our investigation of EM counterparts is organized as follows.  We
begin by comparing various types of EM counterparts, each illustrated
by the schematic diagram in Figure~\ref{fig:cartoon}.  The first is an
SGRB, powered by accretion following the merger ($\S\ref{sec:GRB}$).
Even if no SGRB is produced or detected, the merger may still be
accompanied by relativistic ejecta, which will power non-thermal
afterglow emission as it interacts with the surrounding medium.  In
$\S\ref{sec:ag}$ we explore the properties of such ``orphan
afterglows'' from bursts with jets nearly aligned towards Earth
(optical afterglows; $\S\ref{sec:oa}$) and for larger viewing angles
(late radio afterglows; $\S\ref{sec:ra}$).  We constrain our models
using the existing observations of SGRB afterglows, coupled with
off-axis afterglow models.  We also provide a realistic assessment of
the required observing time and achievable depths in the optical and
radio bands.  In $\S\ref{sec:kilonova}$ we consider isotropic optical
transients powered by the radioactive decay of heavy elements
synthesized in the ejecta (``kilonovae'').  In $\S\ref{sec:compare}$
we compare and contrast the potential counterparts in the context of
our four Cardinal Virtues.  Although some of these counterparts have
been discussed previously in the literature, we examine them together
to better highlight their relative strengths and weaknesses.  Drawing
on the properties of the various counterparts, in \S\ref{sec:strategy}
we make specific recommendations for optimizing the follow-up with
$\gamma$-ray satellites, wide-field optical telescopes (PTF,
Pan-STARRS, LSST), and radio telescopes (EVLA, ASKAP).  We summarize
our conclusions in $\S\ref{sec:conclusion}$.

\section{Short-Duration Gamma-Ray Bursts}
\label{sec:GRB}

The most commonly discussed EM counterpart of NS-NS/NS-BH mergers is
an SGRB, powered by accretion onto the central compact object (e.g.,
\citealt{Paczynski86,Eichler+89,Narayan+92,Rezzolla+11}).  The {\it
Swift} satellite, and rapid follow-up observations with ground-based
telescopes, have revolutionized our understanding of SGRBs by
detecting and localizing a significant number of their afterglows for
the first time (e.g., \citealt{Berger+05,Fox+05,Hjorth+05,Bloom+06}).
This has enabled the discovery that SGRBs originate from more evolved
stellar populations than those of long-duration GRBs, consistent with
an origin associated with NS-NS mergers
\citep{Berger+05,Bloom+06,Leibler&Berger10,Berger11,Fong+11}.  The study
of SGRB afterglows has also established a scale for the energy release
and circumburst density that are lower than for long GRBs, with
$E\lesssim 10^{51}$ erg and $n\lesssim 0.1$ cm$^{-3}$
\citep{Berger+05,Soderberg+06,Berger+07}.  These observations have
also provided evidence for collimation in at least one case
(GRB\,051221A), with a jet half-opening angle of $\theta_j\approx
0.12$ \citep{Burrows+06,Soderberg+06}, and upper or lower limits in
additional cases \citep{Fox+05,Grupe+06,Berger07}, overall suggestive
of wider opening angles than for long GRBs.

Despite this progress, it is not yet established that all SGRBs are
uniquely associated with NS-NS/NS-BH mergers (e.g.,
\citealt{Hurley+05,Metzger+08}), nor that all mergers lead to an
energetic GRB.  The energy of the GRB jet, for instance, may depend
sensitively on the mass of the remnant accretion disk, which from
numerical simulations appears to vary by orders of magnitude ($\sim
10^{-3}-0.1$ M$_\odot$), depending on the properties of the binary and
the high-density equation of state
\citep{Ruffert+97,Janka+99,Lee01,Rosswog+03,Shibata&Taniguchi08,Duez+09,Chawla+10}.

\begin{figure}
\centerline{\psfig{file=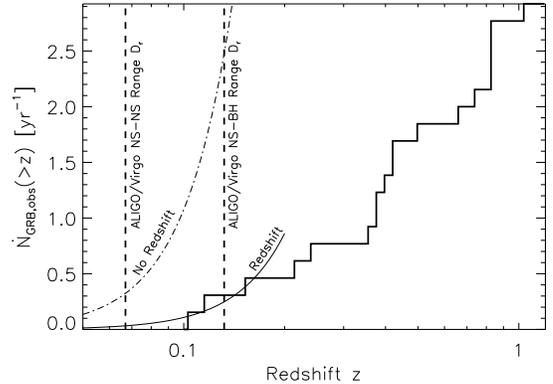,width=3.2in}}
\caption[]{Cumulative detection rate of SGRBs with measured redshifts
$>z$ ({\it thick solid line}), calculated using 19 (mostly {\it
Swift}) SGRBs (e.g., \citealt{Berger11}).  Dashed vertical lines mark
the estimated sensitivity range of ALIGO/Virgo to NS-NS and NS-BH
mergers, respectively, including a boost due to the face-on binary
orientation.  The thin solid line shows an approximate fit to
$\dot{N}_{\rm GRB,obs}(>z)$ at low redshift.  The dot-dashed line
shows an estimate of the {\it total} SGRB detection rate (with or
without redshift information) by an all-sky $\gamma$-ray telescope
with a sensitivity similar to {\it Fermi}/GBM.}
\label{fig:redshift}
\end{figure}

Although SGRBs are bright, they occur relatively rarely within the
range of ALIGO/Virgo.  To illustrate this point, in
Figure~\ref{fig:redshift} we plot the cumulative rate at which SGRBs
are currently detected above a redshift $z$, $\dot{N}_{\rm GRB,obs}
(>z)$.  This distribution includes 19 SGRBs with well-determined
redshifts, obtained from host galaxy associations (e.g.,
\citealt{Berger09}).  Since its launch in late 2004 {\it Swift} has
detected SGRBs at a rate of $\sim 10$ yr$^{-1}$, of which $\sim 1/3$
have measured redshifts.  Shown for comparison are the sensitivity
ranges $D_{\rm r} \approx 1.5\times 196[410] \approx 295[615]$ Mpc for
detection of NS-NS[NS-BH] mergers by
ALIGO/Virgo\footnotemark\footnotetext{Throughout this paper we adopt
the fiducial values for $D_{\rm r}\approx 200$ Mpc from
\citet{Abadie+10}, who define detections as events with
signal-to-noise ratio (SNR) of 8 in a single detector, assuming NS/BH
masses of 1.4/10$M_{\odot}$.  This choice is conservative because for
a network of $N$ detectors, the sensitivity range at fixed SNR
increases $D_{\rm r}\propto N^{1/2}$.  On the other hand, the real
detection range of a network depends on the data quality (e.g.,
Gaussianity and stationarity) and detection pipeline.  Once a value
for $D_{\rm r}$ is chosen, all of the results presented in this paper
may be rescaled accordingly.}, where the factor of $\approx 1.5$
(included only in this section and $\S\ref{sec:oa}$) accounts for the
stronger GW signal from face-on mergers, which characterize the
geometry of GRB jets (e.g., \citealt{Kochanek&Piran93,Schutz11}).

Figure~\ref{fig:redshift} illustrates the striking fact that no SGRBs
with known redshifts have yet occurred within the ALIGO/Virgo range for
NS-NS mergers, while only two SGRBs (061201 and 080905) have occurred
within the NS-BH range.  Though selection effects and low-number
statistics undoubtedly distort the true redshift distribution from
that shown in Figure~\ref{fig:redshift}, at low redshift the
distribution should nevertheless scale as $\dot{N}_{\rm
GRB,obs}\propto z^{3}$.  By fitting the lowest redshift bins to a
distribution of this form, we find that $\lesssim 0.03(0.3)$ SGRBs per
year are currently being localized by {\it Swift} within the
ALIGO/Virgo range for NS-NS(NS-BH)
mergers\footnotemark\footnotetext{The sensitivity range for a GW
detection may be increased somewhat if the search is restricted to the
time interval and sky position of the SGRB in the case of a
$\gamma$-ray triggered search \citep{Kochanek&Piran93}, but this does
not alter our conclusion that SGRBs are a rare occurrence in the range
of ALIGO/Virgo.}.  Thus, even assuming that {\it Swift} (or a mission
with similar capabilities) operates simultaneously with ALIGO/Virgo,
SGRBs are clearly not ideal counterparts to localize a large number of
mergers.  Obtaining a single GW redshift in this fashion could require
a decade of observations.

Localization is of course only one desirable virtue of an EM
counterpart.  Due to the short duration of both SGRBs and the GW
signal, and the short expected delay ($\lesssim\,$seconds) between
them, a time coincidence between these events is sufficient to enable
a statistically confident association.  Even if the redshift cannot be
obtained, a coincident detection will still confirm the astrophysical
nature of the GW signal, prove the connection between SGRBs and
NS-NS/NS-BH mergers, and allow studies of the dependence of the binary
inclination on the properties of the GRB jet (e.g.,
\citealt{Kochanek&Piran93}).  Coincidence searches for GW bursts using
the time and sky coordinates of detected SGRBs were already conducted
during previous LIGO/Virgo Science Runs (e.g.,
\citealt{Abadie+10b,Abbott+10})

To estimate how long ALIGO/Virgo must operate before a connection
between SGRBs and NS-NS/NS-BH mergers can be tested, we also plot in
Figure~\ref{fig:redshift} an estimate of the low-redshift
distribution, but including {\it all} detectable SGRBs (with or
without redshift information), which we estimate by multiplying the
``with redshift'' distribution by a factor $\approx 10$.  This factor
accounts for the higher rate, $\sim 20$ yr$^{-1}$, that {\it Fermi's}
Gamma-Ray Burst Monitor (GBM) detects SGRBs, as compared to the rate
with redshift from {\it Swift} (correcting also for the GBM field of
view, which covers only $\sim 60$ per cent of the sky).  This estimate
illustrates that a few {\it Fermi} bursts over the past few years
probably occurred within the ALIGO/Virgo volume.  Thus, an all-sky
$\gamma$-ray monitor with a sensitivity similar to {\it Fermi}/GBM
could test whether SGRBs originate from NS-NS/NS-BH mergers within
just a few years after ALIGO/Virgo reaches full sensitivity, even if
it does not lead to a significant improvement in the sky
localizations.

One issue raised by the above analysis is that the observed SGRB rate
within the ALIGO/Virgo volume, even when corrected for partial sky
coverage, is much lower than the best-bet NS-NS merger rate of $\sim
40$ yr$^{-1}$.  \citet{Nakar+06} estimate that the local volumetric
SGRB rate is $\gtrsim 10$ Gpc$^{-3}$ yr$^{-1}$, which corresponds to
an all-sky rate of $\dot{N}_{\rm GRB,all-sky}\sim 0.3$ yr$^{-1}$ at a
distance of $\lesssim D_{\rm r,NS-NS}\approx 200$ Mpc (cf.,
\citealt{Guetta&Piran05}), consistent with our estimates in
Figure~\ref{fig:redshift} and still two orders of magnitude below
$\sim 40$ yr$^{-1}$.  Reconciling this remaining discrepancy requires
either that the true merger rate is lower than the best-bet rate; that
all mergers are not accompanied by a bright SGRB; or that the
$\gamma$-ray emission is beamed (e.g.,~\citealt{Rosswog&Ramirez-Ruiz02,Aloy+05}).

Expanding on this final possibility, if the typical SGRB jet has a
half-opening angle $\theta_j\lesssim \pi/2$, then only a fraction
$f_{b,\gamma}\approx 1-\cos\theta_j\approx \theta_{\rm j}^{2}/2\ll 1$
of viewers with observing angles $\theta_{\rm obs}\lesssim \theta_j$
will detect a bright SGRB.  For all other observers (the majority of
cases) the prompt emission is much dimmer due to relativistic beaming.
Reconciling the ``observed'' and best-bet rate by beaming alone thus
requires $f_{b,\gamma}\sim 0.01$, or $\theta_j\sim 0.12$, similar to
the opening angle inferred for GRB\,051221A
\citep{Burrows+06,Soderberg+06}.

A mystery associated with SGRBs is that $\sim 1/4-1/2$ are followed by
variable X-ray emission with a fluence comparable to, or in excess of
the initial burst (e.g., \citealt{Norris&Bonnell06};
\citealt{Perley+09}).  Although the origin of this extended emission
is still debated, one explanation is that it results from ongoing
energy output from a highly magnetized neutron star, which survives
the NS-NS merger (\citealt{Metzger+08}; \citealt{Bucciantini+11}).
Regardless of its origin, if some mergers are indeed accompanied by
extended X-ray emission, this provides an additional potential EM
counterpart, especially if the X-ray emission is more isotropic than
the SGRB itself (as predicted by several models:
\citealt{MacFadyen+05,Metzger+08,Barkov&Pozanenko11,Bucciantini+11}).
Considering alternative prompt counterparts is germane because the
lifetime of {\it Swift} and {\it Fermi} are uncertain, while the next
generation of proposed high-energy transient satellites (e.g., Janus,
Lobster) are most sensitive at soft X-ray (rather than $\gamma-$ray)
energies, which could reduce their sensitivity to detecting the prompt
SGRB phase.  The difficulty of relying on this extended X-ray signal
is twofold: (i) separating such cases from soft long GRBs; and (ii)
these events represent only a fraction of all SGRBs.

\section{Afterglows}
\label{sec:ag}

Even in the absence of an SGRB, an orphan afterglow may provide a
bright electromagnetic link to a GW trigger (e.g.,~
\citealt{Coward+11,Nakar&Piran11}).  The orphan afterglow can be
on-axis if the $\gamma$-ray emission was missed due to incomplete sky
coverage by $\gamma$-ray satellites, or it can be off-axis if the
relativistic jet was initially pointed away from our line of sight.
For off-axis observers the afterglow emission peaks at a later time
and at a lower brightness level than for on-axis observers, making the
detection of a counterpart more challenging.  However, a higher
fraction of events, $\propto\theta_{\rm obs}^{2}$, occur at larger
angles, with the total fraction of detectable counterparts depending
on the largest viewing angle at which emission is still detectable.
On a timescale of $\sim\,$days after the merger, the afterglow
emission is still partially beamed and peaks at optical wavelengths
($\S\ref{sec:oa}$).  At later times, weeks-months, the emission is
mostly isotropic and peaks at radio wavelengths, once the jet
decelerated to mildly relativistic velocities, $\beta \lesssim 1$
($\S\ref{sec:ra}$).

\subsection{Optical Afterglow}
\label{sec:oa}

\begin{figure*}
\centerline{\psfig{file=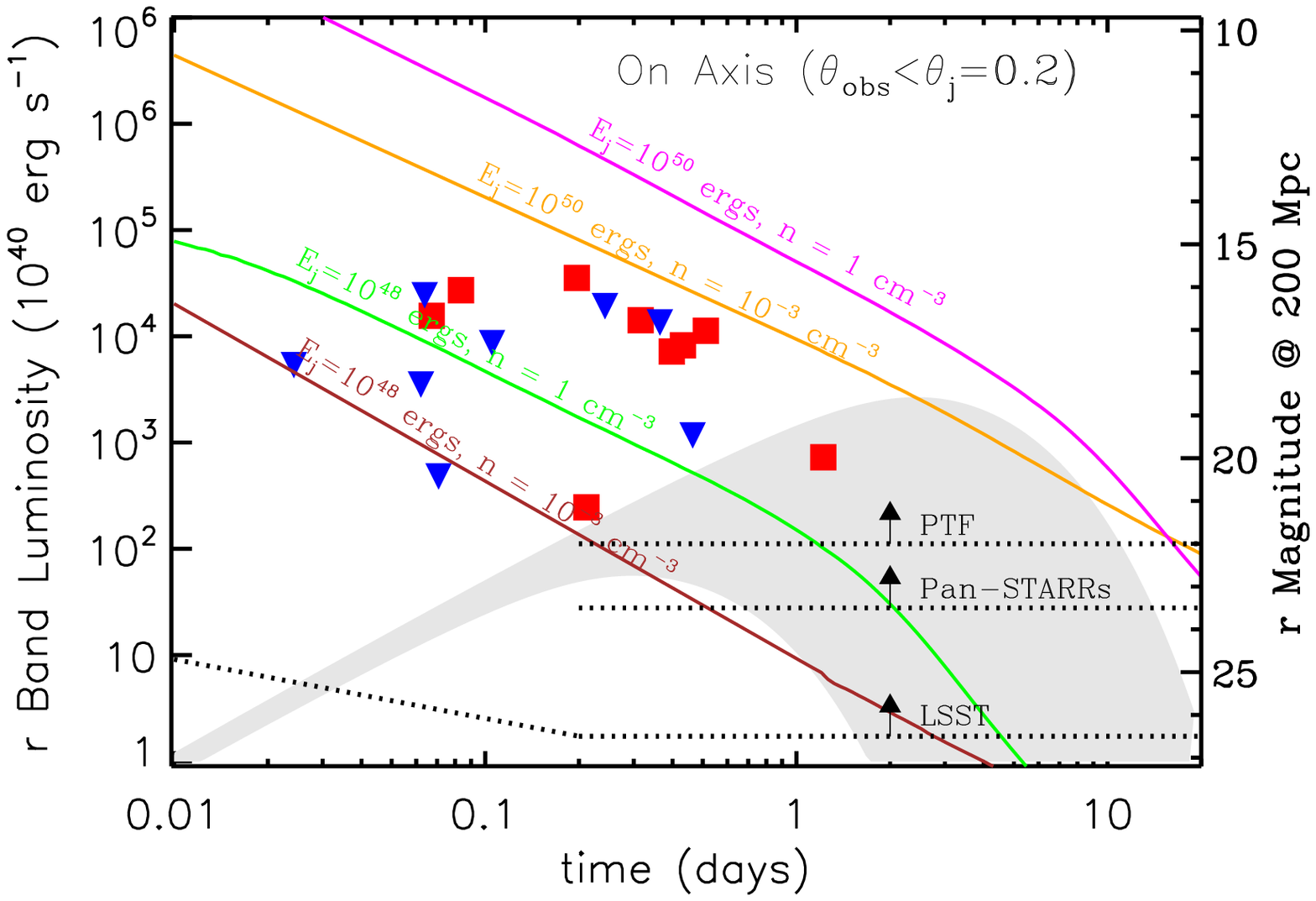,width=6.4in}}
\caption[]{Optical rest-frame luminosity of existing SGRB afterglows
(detections={\it red squares}; upper limits={\it blue triangles};
\citealt{Berger10,Fong+11}).  Solid lines are afterglow models from
\citet{VanEerten&MacFadyen11} (see also \citealt{VanEerten+10}),
calculated for on-axis observers ($\theta_{\rm obs}=\theta_{\rm
j}=0.2$) for a range of jet energies ($E_{\rm j}$) and circumburst
densities ($n$).  The existing afterglows define an upper bound on a
figure of merit, $FOM_{\rm opt,on}\equiv E_{j,50}^{4/3}\,n_0^{1/2}
\lesssim 0.1$.  Also shown are a range of plausible kilonova models
({\it gray shading}).  The $5\sigma$ limiting magnitudes of various
wide-field telescopes are marked by dashed lines; for PTF and
Pan-STARRS we assume a maximum of 0.5 hr per pointing to cover a
typical GW error region with a 1-day cadence, while for LSST we show
both the normal survey depth and the depth for 0.5 hr exposures.}
\label{fig:onaxis}
\end{figure*}

To gain insight into the afterglow emission that may accompany a
merger event (on- or off-axis), we use existing observations of SGRB
optical afterglows discovered in rapid follow-up observations.
Figure~\ref{fig:onaxis} shows optical detections and upper limits from
the compilation of \citet{Berger10} and \citet{Fong+11}, expressed in
luminosity and in apparent magnitude for a source at a distance of
$200$ Mpc.  Shown for comparison are on-axis SGRB afterglow models
with $\theta_{\rm obs}\approx \theta_j$ from
\citet{VanEerten&MacFadyen11}\footnotemark\footnotetext{\url{http://cosmo.nyu.edu/afterglowlibrary/}}
(see also \citealt{VanEerten+10}) that span the range of detected
afterglows.  Also shown are the sensitivity limits of existing and
planned wide-field survey telescopes (PTF, Pan-STARRS, LSST), taking
into account that about 10 pointings will be required to cover a
typical GW error region, leading to at most 0.5 hr per pointing to
cover the region in a single night; for LSST we also show the expected
depth of the normal survey mode ($r\approx 24.7$ mag), which can cover
a GW error region in only a few minutes.  The expected maximum depth
is about 22 mag for PTF, 23.5 mag for Pan-STARRS, and 26.5 mag for
LSST.  We note that the trade-off between limiting magnitude and
localization area ($A$) is simply $\Delta m\approx -2.5{\rm log}
(A^{1/2})$ such that for a best-case scenario of $A\sim {\rm few}$
deg$^{2}$, these telescopes can achieve a greater depth by about 1.2
mag in a single night.

Figure~\ref{fig:onaxis} demonstrates that for a typical jet
half-opening angle of $\theta_j=0.2$, afterglow models with jet
energies of $E_j\approx 10^{48}-10^{50}$ erg and circumburst densities
of $n\approx 10^{-3}-1$ cm$^{-3}$ are consistent with the range of
observed optical luminosities.  We can define a figure of merit for
the combination of energy and density (e.g., \citealt{gs02}), which
based on the observed on-axis optical afterglow luminosities has an
upper bound of:
\begin{equation} 
FOM_{\rm opt,on}\equiv E_{j,50}^{4/3}\,n_0^{1/2}\lesssim 0.1 
\label{eqn:fomo} 
\end{equation}
and a mean value for the detected sample of $FOM_{\rm opt,on}\sim
0.01$; here $E_{j,50}$ is the jet energy in units of $10^{50}$ erg,
$n_0$ is the circumburst density in units of cm$^{-3}$, and we assume
a typical value of $p=2.5$ for the electron power law distribution.
We note that the sample in Figure~\ref{fig:onaxis} represents all
SGRBs with deep optical searches, and hence also a detected X-ray
afterglow.  Since $\sim 1/4$ of SGRBs lack detected X-ray afterglows,
and not all events with X-ray detections had deep follow-up optical
searches, it is possible that some SGRB optical afterglows are dimmer
than those in Figure~\ref{fig:onaxis}, leading to an even lower mean
value of $FOM_{\rm opt,on}$ than inferred above.  Nevertheless, we
conclude that $\sim 1/2$ of SGRBs within the range of ALIGO/Virgo
(even those missed due to incomplete $\gamma$-ray sky coverage) should
produce optical emission detectable by LSST for at least $\sim 10$ d;
the brightest events should be detectable for a few days even by less
sensitive surveys such as PTF.

\begin{figure}
\centerline{\psfig{file=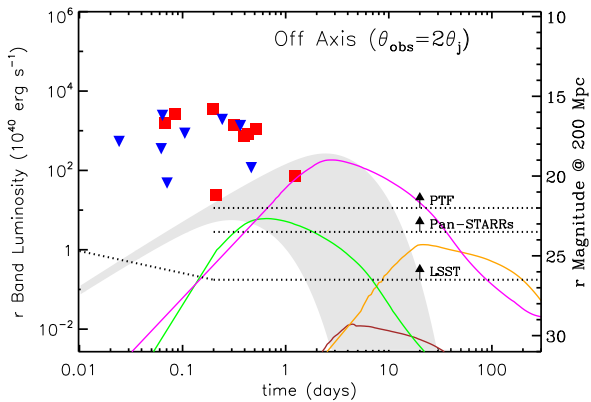,width=3.2in}} 
\caption[]{Same as Figure~\ref{fig:onaxis} but for off-axis observers
with $\theta_{\rm obs}\approx 2\theta_j$.  The kilonova emission
({\it gray shading}) is isotropic and hence remains unchanged for both
on- and off-axis observers.  The range of existing SGRB optical
afterglows, covered by the yellow, green, and brown lines, indicates
that observations with LSST are essential.}
\label{fig:offaxis}
\end{figure}

\begin{figure}
\centerline{\psfig{file=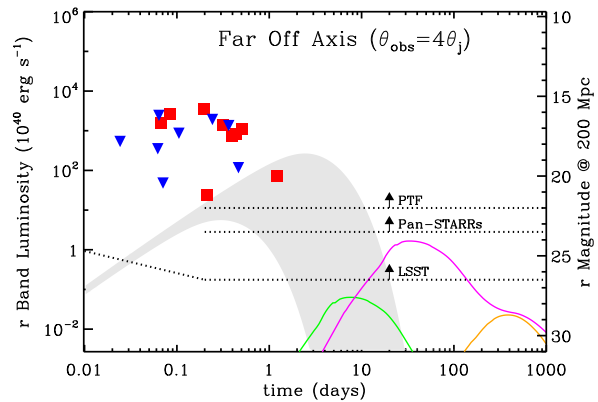,width=3.2in}} 
\caption[]{Same as Figure~\ref{fig:offaxis} but for off-axis observers
with $\theta_{\rm obs}\approx 4\theta_j$.  The kilonova emission
({\it gray shading}) is isotropic and hence remains unchanged for both
on- and off-axis observers.  The range of existing SGRB optical
afterglows, covered by the yellow, green, and brown lines (well below
the limit of the plot), indicates that no existing or future telescope
will be able to detect optical emission at such large off-axis
angles.}
\label{fig:offaxis2}
\end{figure}

Using the range of energies and circumburst densities inferred for
on-axis afterglows, we can now predict the appearance of off-axis
light curves.  Figure~\ref{fig:offaxis} shows the same range of models
from Figure~\ref{fig:onaxis}, but now for an observer angle of
$\theta_{\rm obs}=2\theta_j$.  For the parameters of existing SGRB
afterglows, the peak flux at even larger viewing angles (e.g.,
$\theta_{\rm obs}=4\theta_j$; Figure~\ref{fig:offaxis2}) is too low to
be detected even with LSST.  For the range of applicable models, the
off-axis light curves rise to maximum brightness on a timescale of
$\sim 1-20$ d, with a peak luminosity of $\sim 10^{38}-10^{41}$ erg
s$^{-1}$; this corresponds to a apparent brightness of $\gtrsim 23$
mag at $200$ Mpc.  The models with $n=10^{-3}$ cm$^{-3}$ peak on a
timescale about 7 times longer, and with a luminosity that is about
300 times lower, than those with $n=1$ cm$^{-3}$.  We also note that
the off-axis afterglow light curves in the higher density cases are
qualitatively similar to those of kilonovae, although the latter fade
more rapidly after the peak and have a distinct color evolution
(\S\ref{sec:kilonova}).

\begin{figure*}
\centerline{\psfig{file=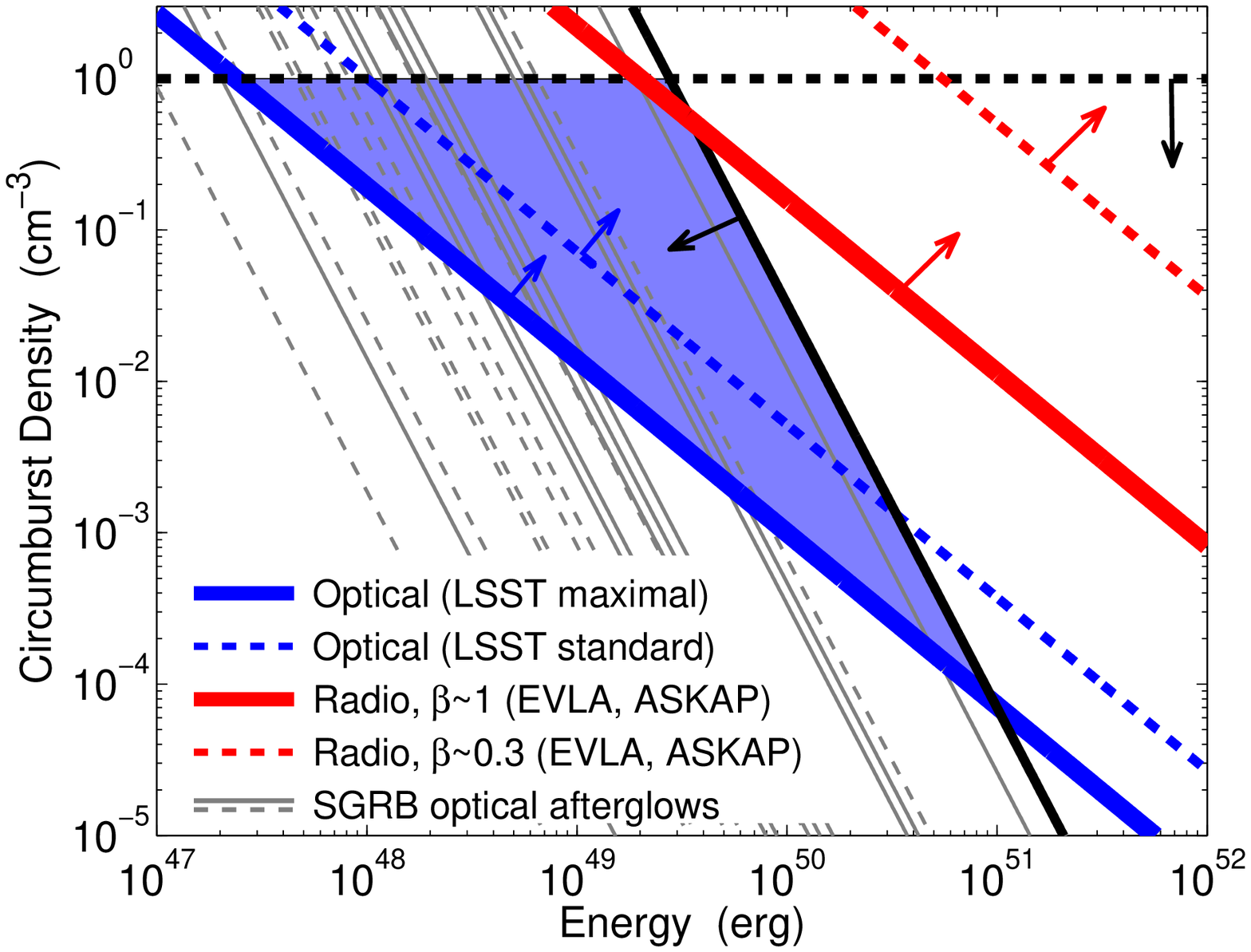,width=6.4in}} 
\caption[]{Phase-space of energy and circumburst density that is
accessible to off-axis afterglow searches in the optical (shaded blue)
and radio (red).  The solid and dashed blue lines mark the lower
bounds for searches with a maximal LSST depth (26.5 mag) and the
standard depth (24.7 mag).  The dashed red line marks the lower bound
for radio emission from ejecta with $\beta\sim 0.3$.  The solid black
line corresponds to the {\it upper} bound defined by existing SGRB
optical afterglows ($FOM_{\rm opt,on}\lesssim 0.1$;
Equation~\ref{eqn:fomo}), while the black dashed line marks the
expected upper bound on the density ($n=1$ cm$^{-3}$) for mergers in
the ISM of a disk galaxy.  Finally, the gray solid (dashed) lines mark
the tracks for existing SGRB optical afterglow detections (limits)
from Figure~\ref{fig:onaxis}.  The existing data suggest that radio
detections are highly unlikely.  On the other hand, the phase-space
accessible to optical searches is populated by at least some of the
existing events.}
\label{fig:fom}
\end{figure*}

For the off-axis light curves we can define a separate figure of merit
(c.f., Equation 11 of \citealt{Nakar&Piran11}):
\begin{equation}
FOM_{\rm opt,off}\equiv E_{j,50}\,n_0^{7/8}.
\label{eqn:fomoff}
\end{equation}
We define detectable cases as those rising by at least an order of
magnitude (2.5 mag) above the LSST maximal depth, corresponding to
$FOM_{\rm opt,off}\gtrsim 0.002$ (Figures~\ref{fig:offaxis} and
\ref{fig:fom}).  With the same criterion for the threshold, a shallow
survey such as PTF will only detect events with $FOM_{\rm opt,off}\sim
1$, beyond the range of existing on-axis SGRB afterglows.  In
Figure~\ref{fig:fom} we plot the detectable region in the $E_j-n$
phase-space for the maximal LSST depth and the LSST normal survey
depth ($FOM_{\rm opt,off}\gtrsim 0.01$).  The allowed phase-space is
bounded by the on-axis figure of merit (Equation~\ref{eqn:fomo}), and
we also introduce an upper density cut-off of $n\lesssim 1$ cm$^{-3}$
as an optimistic density for a merger in the interstellar medium of a
disk galaxy.  Due to the different dependencies of $FOM_{\rm opt,on}$
and $FOM_{\rm opt,off}$ on $E_j$ and $n$, these conditions define a
triangular region of allowed phase-space for detections of off-axis
optical afterglows; in \S\ref{sec:ra} we perform a similar calculation
for off-axis radio afterglows to compare the relative $E_j-n$
phase-space that is probed by each band.  Most importantly, we find
that the tracks for existing SGRB afterglows cross the phase-space
region covered by optical searches.

We explore the detectability of on- and off-axis optical afterglows
more precisely with a Monte Carlo simulation to determine the fraction
of GW events within 200 Mpc that would be detected by an optical
survey with a given limiting magnitude and cadence.  For targeted
follow-up searches we use a 1-day cadence with limiting magnitudes of
22 (PTF), 23.5 (Pan-STARRS), and 26.5 (LSST); we also include a 1-day
and 4-day cadence with the standard LSST depth of 24.7 mag.  The
results, summarized in Table~\ref{tab:afterglow}, show that if the
jet energy and circumburst density are similar to those required to
explain the on-axis SGRB data (Figure~\ref{fig:onaxis}), then events
with $\theta_{\rm obs}\lesssim 2\theta_j$ are sufficiently
bright to be detected in at least 3--5 epochs, given a survey with a
depth similar to the standard LSST survey ($24.7$ mag), but with a
faster cadence of $\sim 1$ d.  Shallower searches are also capable of
detecting energetic afterglows in a few epochs, but this may not be
sufficient for a clear identification.  By contrast, in most cases
events viewed at larger angles ($\theta_{\rm obs}\gtrsim 2\theta_{\rm
j}$) are not detectable, even near peak emission with LSST.  

\begin{figure}
\centerline{\psfig{file=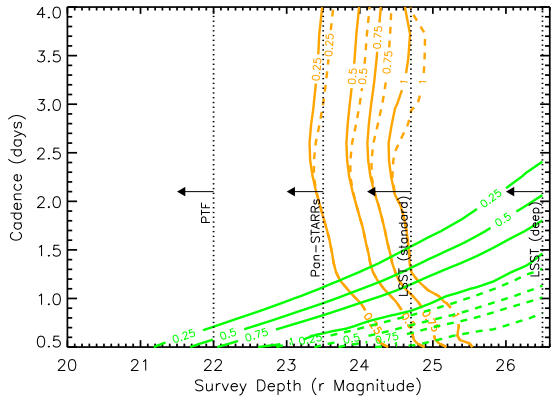,width=3.2in}} 
\caption[]{Fraction of off-axis optical afterglow events detected in 3
({\it solid line}) or 5 ({\it dashed line}) epochs as a function of
the depth and cadence of a search.  The two models shown -- $E_{j}=
10^{50}$ erg, $n = 10^{-3}$ cm$^{-3}$ and $E_{j} = 10^{48}$ erg; $n =
1$ cm$^{-3}$ -- have the same colors as in Figure \ref{fig:onaxis}.}
\label{fig:agfrac}
\end{figure}

The same information is presented graphically in
Figure~\ref{fig:agfrac} where we plot contours of detection fraction
in 3 and 5 epochs as a function of depth and cadence.  We find that in
the case of $E_{j}\sim 10^{50}$ erg, the standard LSST cadence and
depth are sufficient for multiple detections.  However, for lower
energies (which may be typical of most SGRBs), a faster cadence and
greater depth ($\sim 26.5$ mag) are required for multiple detections.
To achieve a detection fraction of $50\%$ in 3(5) epochs for the case
of $\theta_{\rm obs}=2\theta_j$ requires a depth of at least
$23.5$($26$) mag for a 1-day cadence.

Since detectable optical emission is limited to off-axis angles of
$\lesssim 2\theta_j$, we estimate the corresponding fraction of
GW events with potential optical afterglow detections as:
\begin{equation}
f_{\rm opt}\approx \int_{0}^{2\bar{\theta}_{\rm j}}p_{\rm det}d\theta
\approx 6.8\bar{\theta}_{\rm j}^{2}+
\mathcal{O}\left(\bar{\theta_j}^{4} \right),
\label{eq:fopt}
\end{equation}
where $\bar{\theta}_{\rm j}\ll 1$ is the average opening angle, and
\begin{equation}
p_{\rm det}(\theta)\approx 0.152\sin\theta\left(1+6\cos^{2}\theta+
\cos^{4}\theta\right)^{3/2}
\label{eq:pdetect}
\end{equation}
is the detection probability of events with inclination angles between
$\theta$ and $\theta+d\theta$ (e.g., Schutz 2011; their Equation 28).
As noted earlier, since this scenario is nearly face-on, $f_{\rm opt}$
is a factor $\sim 3.4$ higher than the detection fraction $\approx
1-\cos 2\theta_j\approx 2\bar{\theta_j}^{2}$ for isotropic emission.

Equation~\ref{eq:fopt} shows that if the average opening angle is
$\bar{\theta}_j\simeq 0.12$, which is the value inferred for
GRB\,051221A \citep{Burrows+06,Soderberg+06}, as well as the typical
opening angle required to reconcile the observed SGRB rate with the
best-bet NS-NS merger rate (\S\ref{sec:GRB}), then up to $f_{\rm opt}
\sim 0.1$ of GW events will be accompanied by potentially detectable
optical afterglows.  This result is consistent with the rate of a few
afterglows per year inferred by \citet{Coward+11} for their assumed
total ALIGO/Virgo merger rate of $\sim 135$ yr$^{-1}$.  On the other
hand, if $\bar{\theta_j}$ is much larger, $\gtrsim 0.4$ (e.g.,~ as
found for GRB\,050724 by \citealt{Grupe+06}), then $f_{\rm opt}$ is of
order unity, but the overall GW event rate may be lower than the best-bet
ALIGO/Virgo rate.

Beyond considerations of depth and cadence, a unique optical
identification of GW events also requires discrimination between
off-axis afterglows and potential contaminants.  We discuss this issue
in \S\ref{sec:compare}.

\subsection{Radio Afterglow}
\label{sec:ra}

NS-NS/NS-BH mergers may also be accompanied by non-thermal radio
afterglow emission, which can originate either from the
ultra-relativistic jet (as in the case of the optical afterglow), or
from more spherical, sub-relativistic ejecta (\citealt{Nakar&Piran11};
hereafter NP11).  The latter includes matter ejected dynamically
during the merger process (``tidal tails''), or in outflows from the
accretion disk (see Figure~\ref{fig:cartoon}).  Adopting standard
models for synchrotron emission from a relativistic shock, NP11
estimate that the peak radio brightness for these cases is:
\begin{equation}
F_{\rm\nu,p}\approx 40\,E_{j,50}\, n_{0}^{7/8}\,\beta_{0.2}^{11/4}\,
d_{L,200}^{-2}\, \nu_1^{-3/4}\,\,\,{\rm \mu Jy},
\label{eqn:Fp}
\end{equation}
where $\beta_{0.2}=v_{\rm ej}/0.2c$, $\nu_1$ is the observing
frequency in GHz; and $d_{L} = 200 d_{L,200}$ Mpc is the luminosity
distance, again normalized to the ALIGO/Virgo range for NS-NS mergers.
Equation~\ref{eqn:Fp} also assumes characteristic values of $p=2.5$
for the electron distribution power law index, and $\epsilon_{e}=
\epsilon_{B}=0.1$ for the fractions of energy density imparted to
relativistic electrons and magnetic fields, respectively.  The radio
emission peaks at the deceleration time:
\begin{equation} 
t_{\rm dec}\approx 2.6\,E_{j,50}^{1/3}\, n_{0}^{-1/3}\,
\beta_{0.2}^{-5/3}\,\,\,{\rm yr}.
\label{eqn:tdec}
\end{equation} 
The peak brightness depends sensitively on both the properties of the
ejecta ($E$ and $\beta$) and on the circumburst density.  As we
discuss in detail below, the realistic detection threshold for a
convincing detection with the EVLA (even with $\sim 30$ hr per epoch)
is about 0.5 mJy.  This requirement therefore defines a figure of
merit for a radio detection of:
\begin{equation}
FOM_{\rm rad}\equiv E_{j,50}\,n_0^{7/8}\,\beta^{11/4}\gtrsim 0.2.
\label{eqn:fomr}
\end{equation}
With the exception of the velocity parameter, this figure of merit is
identical to the case of off-axis optical afterglows in terms of the
dependence on $E_j$ and $n$.

For quasi-spherical ejecta, a characteristic mass of $M_{\rm ej}\sim
10^{-2}$ M$_\odot$ in tidal tails or disk winds has an energy\footnotetext{The ejecta mass and velocity may be higher in some NS-BH mergers (e.g.,
\citealt{Rosswog05}), especially those that merge on eccentric orbits
in dense stellar clusters (e.g., \citealt{Lee+10,Stephens+11}), but
these are unlikely to represent the typical case.}
$E\approx M_{\rm ej}v_{\rm ej}^{2}/2\sim 10^{50}-10^{51}$ erg for the
expected range of velocities\footnotemark $\beta\sim
0.1-0.3$.  This results in at most $FOM_{\rm rad}\approx 0.4\,
n_0^{7/8}$, requiring $n_0\gtrsim 1$ cm$^{-3}$ for a detection.  For
more typical densities of $\lesssim 0.1$ cm$^{-3}$ associated with
SGRBs \citep{Berger+05,Soderberg+06}, the radio emission from
quasi-spherical ejecta will be essentially undetectable unless the
energy scale is much larger than $\sim 10^{51}$ erg
(Figure~\ref{fig:fom}).

Equations~\ref{eqn:Fp}--\ref{eqn:fomr} can also be applied to the case
of off-axis afterglow emission using $\beta\approx 1$
\citep{Nakar&Piran11} along with values for the jet energy and
circumburst density inferred from the optical afterglow data
($FOM_{\rm opt,on}\lesssim 0.1$; Equation~\ref{eqn:fomo}).  In
Figure~\ref{fig:fom} we plot the region of $E-n$ phase-space that is
accessible to radio detections ($FOM_{\rm rad}\gtrsim 0.2$).  As can
be seen from the Figure, none of the existing SGRB optical afterglow
intersect this region, indicating that radio detections of off-axis
afterglows are likely to be rare despite the overall isotropy of the
signal.

We now address in detail the estimated minimum radio brightness
necessary for a successful detection.  Although faint radio emission
is in principle detectable with a deep integration, a significant
challenge is the small field of view of sensitive instruments such as
the EVLA ($\approx 0.4$ deg$^2$ at 1 GHz), requiring $\sim 100-200$
pointings to cover a typical GW error region of tens of square
degrees.  Targeting individual galaxies within the error region does
not decrease the number of required pointings since there are $\sim
400$ galaxies with $L\gtrsim 0.1\,L^*$ within a typical error region
(to 200 Mpc).  Even with only 10 min per pointing, $\sim 30$ hr per
epoch will be required to cover the full error
region\footnotemark\footnotetext{The typical overhead for phase, flux,
and bandpass calibration with the EVLA is about $25\%$.}, already a
substantial allocation of EVLA time.  Multiple epochs will be required
over a span of weeks to years to detect the rise and decline of the
radio light curve following a GW detection (Equation~\ref{eqn:tdec}),
for a total of about $\sim 300$ hr of EVLA time (i.e., to search the
best-bet rate of $\sim 40$ GW triggers per year will require
essentially $100\%$ of the EVLA time).  Thus, a reasonable exposure
time per pointing is $\lesssim 10$ min, which at 1 GHz corresponds to
a $5\sigma$
limit\footnotemark\footnotetext{\url{https://science.nrao.edu/facilities/evla/calibration-and-tools/exposure}}
of about 0.25 mJy.  Since a convincing detection will require the
brightness to rise to about twice the threshold, the minimum
detectable peak flux is $F_{\rm\nu,p}\approx 0.5$ mJy.  We note that
the threshold may be even higher in the compact EVLA configurations (C
and D) due to substantial source confusion imposed by the large
synthesized beam size ($12-44''$).

Observations at a higher frequency of 5 GHz can in principle provide
better sensitivity (and reduce source confusion problems), but in
reality will actually require even more observing time.  This is
mainly because the field of view at 5 GHz is sufficiently small (0.02
deg$^2$) that a more profitable strategy is to target the $\sim 400$
galaxies with $L\gtrsim 0.1\,L^*$ within a typical GW error region.
Even with only 5 min per pointing this will require about 40 hr per
epoch, with a resulting $5\sigma$ limit of 0.1 mJy.  A convincing
detection will therefore require $F_{\rm\nu,p}\gtrsim 0.2$ mJy, which
given a typical spectrum of $F_{\rm\nu,p}\propto \nu^{-0.75}$ is
equivalent to a limit of $\gtrsim 0.7$ mJy at 1 GHz, worse than the 1
GHz observing strategy, with even more time required per epoch.

Observations with future wide-field radio interferometers (e.g.,~
ASKAP) will cover a typical GW error region with a few pointings,
requiring only a few hours per epoch.  However, these instruments
suffer from poorer angular resolution compared to what is possible
with the EVLA (e.g., ASKAP with $\sim 10''$ resolution).  This will
lead to significant source confusion at the required low flux density
levels.  More critically, radio emission from the host galaxy itself
will present a challenge; at 200 Mpc a star formation rate of only 1
M$_\odot$ yr$^{-1}$ corresponds to a 1 GHz flux density of about 0.6
mJy \citep{yc02}.  At a resolution of $10''$ (10 kpc at 200 Mpc)
galaxies will generally appear as unresolved point sources, and will
prevent the detection of significantly fainter coincident radio
counterparts.  Thus, an instrument like ASKAP will cover a GW error
region faster than the EVLA, but to a similar effective depth limited
by source confusion.

A final complication with radio detections is the long time delay
between a GW trigger and the peak of the putative radio signal, which
could negate a robust association.  For a sub-relativistic counterpart
($\beta\sim 0.2$) with an optimistic density of $n\sim 1$ cm$^{-3}$, a
detection requires $E\gtrsim 10^{51}$ erg (Equation~\ref{eqn:fomr}),
and as a result $t_{\rm dec}\approx 6$ yr, requiring observations for
over a decade.  For the relativistic case ($\beta\approx 1$) with
$n\sim 1$ cm$^{-3}$, the peak time corresponding to a detectable
signal is $t_{\rm dec}\approx 0.1$ yr.  The latter case will require a
$\sim\,{\rm week}$ cadence to robustly sample the light curve,
corresponding to about $15-20\%$ of the EVLA time (with $\sim 30$ hr
per epoch).  The absence of a credible detection will require a
$\sim\,{\rm year}$ cadence to search for a non-relativistic
counterpart.  Of course, with a multi-year timescale the probability
of mis-identification with an unrelated radio transient becomes
larger.

Despite the various difficulties outlined above, a clear advantage of
radio searches is the lower number of contaminating sources compared
to the optical band.  As discussed in NP11, confusion with AGN radio
variability can be reduced by requiring an offset from the center of
the host galaxy, although this may be difficult with an angular
resolution of $\gtrsim 10''$ (EVLA in its compact configurations and
ASKAP).  Similarly, while some normal Type Ib/c supernovae have
similar radio light curves to those expected for NS-NS mergers (since
they produce ejecta with $\beta\sim 0.3$), they are generally less
energetic, with only $\sim 10^{47}-10^{48}$ erg coupled to the fast
ejecta \citep{bkc02,bkf+03}.  These events will also be accompanied by
optical supernova emission on a similar timescale, providing an
additional source of discrimination.  Finally, relativistic Type Ib/c
supernovae (with or without an associated GRB) have $\sim
10^{49}-10^{50}$ erg coupled to their fast ejecta
\citep{kfw+98,scp+10}, but these are also accompanied by a bright
optical supernovae.

To conclude, the utility of radio emission as an EM counterpart is
particularly sensitive to the typical energy and circumburst density.
In the case of off-axis afterglows, detections require a high energy
and density that exceed those of known SGRB afterglows
(Figure~\ref{fig:fom}).  In the non-relativistic case, even higher
energy and/or density are required, such that for an expected upper
bound of $n\lesssim 1$ cm$^{-3}$ the required energy is $E\gtrsim
10^{51}$ erg.  The required telescope time for an effective search is
hundreds of hours (EVLA), with perhaps only tens of hours using future
wide-field instruments (e.g.,~ ASKAP).  The time delays range from
months to years, which may complicate a robust association.  The key
advantages are the spherical geometry at $t\gtrsim t_{\rm dec}$ and
the smaller number of contaminating sources compared to the optical
band.

\section{Kilonova}
\label{sec:kilonova}

The detectability of SGRBs and their afterglows is sensitive to
uncertainties in the degree of relativistic beaming and, in the case
of afterglows, the properties of the circumburst environment.  Of
course, it is also possible that not all NS-NS mergers produce SGRBs.
However, independent of this association, the mergers are expected to
be accompanied by isotropic thermal emission, powered by the
radioactive decay of heavy elements in the merger ejecta
(\citealt{Li&Paczynski98}; hereafter LP98;
\citealt{Kulkarni05,Rosswog05,Metzger+10,Roberts+11,Goriely+11}).  Unlike Type
Ia supernovae, which are powered by the decay of $^{56}$Ni and
$^{56}$Co, the ejecta from NS-NS mergers is primarily neutron-rich
(electron fraction $Y_{e}\ll 0.5$) and thus produce little nickel.
Instead, heavier radioactive elements (mass number $A\gtrsim 130$) are
expected to form as neutrons capture onto nuclei ($r$-process
nucleosynthesis) after the ejecta decompresses from nuclear densities
(e.g.,~ \citealt{Lattimer&Schramm74,Eichler+89,Freiburghaus+99}).
Although the $r-$process itself lasts at most a few seconds, these
newly-synthesized elements undergo nuclear fission and beta decays on
much longer timescales.  The resulting energy release will power
bright emission once the ejecta expands sufficiently that photons can
escape.

Neutron-rich material is expected to be ejected both dynamically
during the final coalescence (e.g., \citealt{Rosswog+99}) and by
outflows from the accretion disk at later times (e.g.,
\citealt{Metzger+08b,Metzger+09,Dessart+09,Lee+09}; Figure~\ref{fig:cartoon}).
Depending on the properties of the merging binary, expected values for
the ejecta mass and velocity are in the range $M_{\rm ej}\sim 10^{-3}-
0.1$ M$_{\odot}$ and $\beta \approx 0.1-0.3$, respectively
(e.g.~\citealt{Rosswog+99}; \citealt{Rosswog05}).  The resulting
emission peaks when photons are able to diffuse through the ejecta on
the expansion timescale \citep{Arnett82}; the low ejected mass thus
results in a somewhat dimmer and faster evolving light curve than a
normal supernova, lasting days instead of weeks.

\citet{Metzger+10} use a nuclear physics reaction network to calculate
the radioactive heating of the ejecta from NS mergers, and a radiative
transfer code to model the light curve and color evolution.  For
typical values of $M_{\rm ej}=10^{-2}$ M$_{\odot}$ and $\beta=0.1$,
they find that the transient peaks at an absolute visual magnitude of
$M_{V}\simeq -15$ on a timescale of $\sim 1$ d; because this is
approximately one thousand times brighter than novae (yet dimmer than
{\it super-}novae) they dub these events
kilonovae\footnotemark\footnotetext{The terms {\it mini-supernovae}
(LP98) and {\it macro-novae} \citep{Kulkarni05} are also sometimes
applied.}.

Although the calculations of \citet{Metzger+10} include full radiative
transfer, they show that the kilonova light curve is well-approximated
using a simple one-zone model (LP98), provided that one adopts a value
of $f_{\rm nuc}\approx 3\times 10^{-6}$ for the dimensionless
parameter quantifying the amount of nuclear heating on a timescale of
$\sim 1$ d (LP98).  Similar results for the radioactive heating were
found recently by \citet{Roberts+11} and \citet{Goriely+11}, despite
somewhat different assumptions about the geometric structure and
thermodynamics of the ejecta.

In Figures~\ref{fig:onaxis} and \ref{fig:offaxis} we plot a range of
kilonova models that span the expected range of ejecta mass and
velocity, allowing also for realistic theoretical uncertainties in the
value of $f_{\rm nuc}$ and the opacity of pure $r$-process ejecta.
The resulting kilonova emission peaks on a timescale of $\sim 0.5-5$
d, with an optical luminosity in the range $\sim 10^{41}-10^{42.5}$
erg s$^{-1}$, corresponding to $\approx 19-22.5$ mag at the edge of
the ALIGO/Virgo volume.  Following the peak, the kilonova luminosity
declines as $L_\nu\propto t^{-\alpha}$, with $\alpha\approx 1-1.4$,
due to the declining radioactive power; the actual light curve may
decline even faster once $\gamma$-rays or $\beta$-decay leptons freely
escape the ejecta without depositing their energy.

An important characteristic of kilonovae are their relatively unique
spectra, which can serve to distinguish these events from other
astrophysical transients.  Overall, the kilonova spectrum is predicted
to be quasi-thermal with $T\approx 10^{4}$ K (although line blanketing
in the UV may substantially redden the color temperature).  Near peak,
doppler broadening caused by the high ejecta velocity will smear out
individual spectral features and the overall continuum will be smooth.
Following the peak, however, the photosphere will recede deeper into
the ejecta, where the velocity is lower.  Individual spectral lines
from resonant transitions may then become apparent.  Since the ejecta
are composed entirely of exotic heavy nuclei, the dominant spectral
features may not resemble those of any supernova detected to date.
Detailed predictions of kilonova spectra are unfortunately impossible
because laboratory data on the spectral lines of $r$-process elements
are currently sparse (e.g.,~\citealt{Lawler+09}); the closest known
analog to a pure $r$-process photosphere are ultra metal-poor stars in
the Galactic halo (e.g.,~ \citealt{Sneden+03}).

\begin{figure}
\centerline{\psfig{file=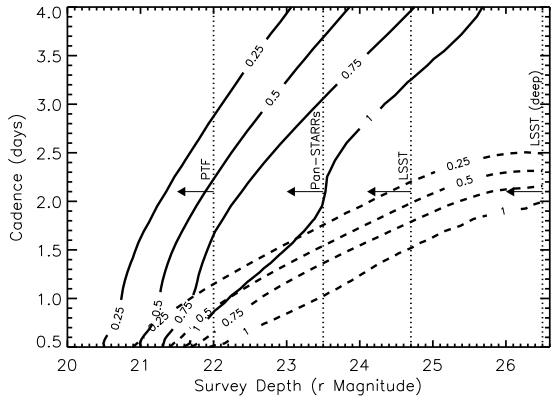,width=3.2in}} 
\caption[]{Fraction of kilonova events detected in 3 ({\it solid
line}) or 5 ({\it dashed line}) epochs as a function of the depth and
cadence of a search.  We have adopted a representative kilonova model
with $M_{\rm ej} = 10^{-2}$ M$_\odot$, $\beta = 0.1$ c, and $f_{\rm
nuc} = 3\times 10^{-6}$.}
\label{fig:LPfrac}
\end{figure}

To assess the detectability of kilonovae we carry out a Monte Carlo
simulation for an optical surveys with a range of limiting magnitudes
and cadences (Table~\ref{tab:kilonova}).  We find that at the depth
and cadence of the normal LSST survey ($r\approx 24.7$ mag, $\Delta
t=4$ d), essentially no kilonovae will be detected in 5 epochs, unless
$M_{\rm ej}\sim 0.1$ M$_\odot$; about 3/4 of all events will be
detected in 3 epochs with the normal LSST survey if $M_{\rm ej}\sim
10^{-2}$ M$_\odot$.  To detect events with $M_{\rm ej}\sim 10^{-2}$
M$_\odot$ in 5 epochs requires a 1-day cadence, preferably with
telescopes capable of reaching $\gtrsim 23$ mag (e.g.,~ Pan-STARRS,
LSST).  Finally, for $M_{\rm ej}\sim 10^{-3}$ M$_\odot$, no existing
or planned telescope will provide 5 detections, but LSST with a 1-day
cadence is likely to provide $3-4$ detections.  These results are also
summarized in Figure~\ref{fig:LPfrac}, where we plot contours for the
fraction of kilonovae detected in 3 and 5 epochs as a function of
limiting magnitude and cadence, assuming typical values of $M_{\rm ej}
= 10^{-2}$ M$_{\odot}$, $\beta = 0.1$, and $f_{\rm nuc} = 3\times
10^{-6}$.  The plot demonstrates that to achieve $50\%$ completeness
in 3(5) epochs given a cadence of $\sim 1$ d requires a limiting
magnitude of $\gtrsim 21$($22.5$) mag.  We discuss potential
contamination from other optical transients in \S\ref{sec:compare}.

\section{Summary: what is the most promising EM counterpart?}
\label{sec:compare}

\begin{deluxetable*}{lccccc}
\tabletypesize{\scriptsize}
\tablecolumns{6}
\tabcolsep0.05in\footnotesize
\tablewidth{0pc}
\tablecaption{Comparison of Electromagnetic Counterparts
\label{tab:compare}}
\tablehead {
\colhead{Counterpart}           &
\colhead{Detection Efficiency}  &
\colhead{Depends on}            &
\colhead{Virtues}               &
\colhead{Follow-up}             \\
\colhead{}                      &
\colhead{Fraction }             &
\colhead{Density?}              &
\colhead{Satisfied}             &
\colhead{Instruments}           
}
\startdata
Short GRB ($\S\ref{sec:GRB}$)               & $\sim 3.4\times f_{b,\gamma}\times$ FOV$_{\gamma}\,^{a}$                                     & no  & 1,3            & {\it Fermi}/GBM  \\
Orphan optical afterglow ($\S\ref{sec:oa}$) & $\sim 7\bar{\theta}_{\rm j}^{2}\times \mathcal{F}_{\rm opt}(E_{\rm j},n)\,^{b}\lesssim 0.1$  & yes & 1,3(?),4       & Pan-STARRS, LSST \\
Orphan radio afterglow ($\S\ref{sec:ra}$)   & $\sim 1\times\mathcal{F}_{\rm rad}(E_{\rm j},n)\,^{c}$                                       & yes & 1,3,4(?)       & EVLA, ASKAP \\
Non-relativistic radio ($\S\ref{sec:ra}$)   & ?; only if $E\gtrsim 10^{51}$ erg                                                            & yes & 1(?),3(?),4(?) & EVLA, ASKAP \\
Kilonova ($\S\ref{sec:kilonova}$)           & $\sim 1\,^{d}$                                                                               & no  & 1,2,3(?),4     & PTF, Pan-STARRS, LSST 
\enddata
\tablecomments{$^{a}$ Field of view of gamma-ray telescope as a
fraction of $4\pi$ steradian; $^b$ fraction of mergers accompanied by
a sufficiently energetic jet and dense circumburst medium for an
optical detection (related to $FOM_{\rm opt,off}\equiv
E_{j,50}\,n_0^{7/8} \gtrsim 0.002$; Equation~\ref{eqn:fomoff}); $^c$
fraction of mergers accompanied by a sufficiently energetic jet and
dense circumburst medium for a radio detection (related to $FOM_{\rm
rad}\equiv E_{j,50}\,n_0^{7/8}\gtrsim 0.2$; Equation~\ref{eqn:fomr});
$^d$ assuming that a telescope similar to LSST covers the GW sky error
region with a cadence of $\sim 1$ day.}
\end{deluxetable*}

We now bring together our conclusions from the previous sections to
address the question of the most promising EM counterpart.  A summary
of the expected detection fractions, dependence on density, and the Cardinal Virtues satisfied for each EM
counterpart is provided in Table~\ref{tab:compare}.  We first discuss
the case in which all NS-NS mergers are accompanied by SGRBs, and
hence by on-axis $\gamma$-ray/afterglow emission or by off-axis
afterglow emission; we then turn to a discussion of the
kilonova-dominated case.

\subsection{Gamma-Rays}

Short GRBs are easily detectable within the ALIGO/Virgo volume with
current $\gamma$-ray satellites in cases when $\theta_{\rm obs}
\lesssim \theta_j$; they therefore satisfy Virtue \#1.  Although this
configuration applies to only a small fraction of all mergers (and
therefore violates Virtue \#2), the SGRB rate within the ALIGO/Virgo
volume (enhanced by a factor of 3.4 for face-on mergers) is
sufficiently high that $\sim 1$ coincident event should occur per year
(Figure~\ref{fig:redshift}).  SGRBs thus represent an ideal
counterpart to confirm the cosmic origin of at least some GW events,
and to test whether SGRBs in fact accompany NS-NS/NS-BH mergers.  Such
an association is critical since it will help to justify the expensive
search for orphan afterglows in the optical and radio bands.  SGRBs
also suffer from little contamination and therefore satisfy Virtue
\#3.  It is therefore critical that a sensitive $\gamma$-ray satellite
be in operation during the ALIGO/Virgo era.  Fine positional accuracy
for an SGRB detection (e.g., {\it Swift}) is less critical than
all-sky coverage (e.g., {\it Fermi}/GBM) since the temporal
association alone within the large error region of a GW source would
suffice to determine an association.  A $\sim\,$arcsecond position
(satisfying Virtue \#4) could then be achieved from the expected
on-axis optical afterglow or a kilonova, which will be brighter than
$\sim 22$ mag and hence easily detectable with wide-field telescopes.
Thus, SGRBs satisfy 3 out of the 4 virtues for a promising EM
counterpart.

\subsection{Off-Axis Optical and Radio Afterglows}

In the absence of $\gamma$-ray emission, orphan afterglow emission
(both optical and radio) is the most promising counterpart if the
typical jet energy and circumburst density lie near the upper end
estimated from current SGRB observations: $E_{j,50}^{4/3}\,n_0^{1/2}
\sim 0.1$. For off-axis optical afterglows the detectability limit
with LSST is $FOM_{\rm opt,off}\equiv E_{j,50}\,n_0^{7/8}\gtrsim
0.002$, as long as $\theta_{\rm obs}\lesssim 2\theta_j$.  Thus,
optical afterglows satisfy Virtue \#1, but violate Virtue \#2 since a
fraction of at most $\sim 7\bar{\theta}_{\rm j}^{2}\sim 0.1$ would be
detectable.  For radio afterglows the detectability limit with EVLA or
a future instrument like ASKAP is $FOM_{\rm rad}\equiv E_{j,50}\,
n_0^{7/8}\gtrsim 0.2$ independent of viewing angle (i.e., they satisfy
Virtue \#2).  However, as a result of the limited range of $E-n$
phase-space probed by radio observations, they violate Virtue \#1.
Indeed, using the tracks for existing SGRB optical afterglows in the
$E-n$ phase-space, we find that none cross the portion accessible to
radio searches (Figure~\ref{fig:fom}).  It is therefore possible that
despite the relative isotropy of the radio emission, existing and
planned instruments are simply not sensitive enough to detect the
emission for typical SGRB parameters.  On the other hand, about half
of all existing optical afterglows will be detectable to the depth of
LSST (Figure~\ref{fig:fom}), but only with a viewing angle up to $\sim
2\theta_j$, indicating an expected optical detection fraction of
$\lesssim 5\%$.

\begin{figure*}
\begin{center}
\includegraphics[angle=0,width=3.3in]{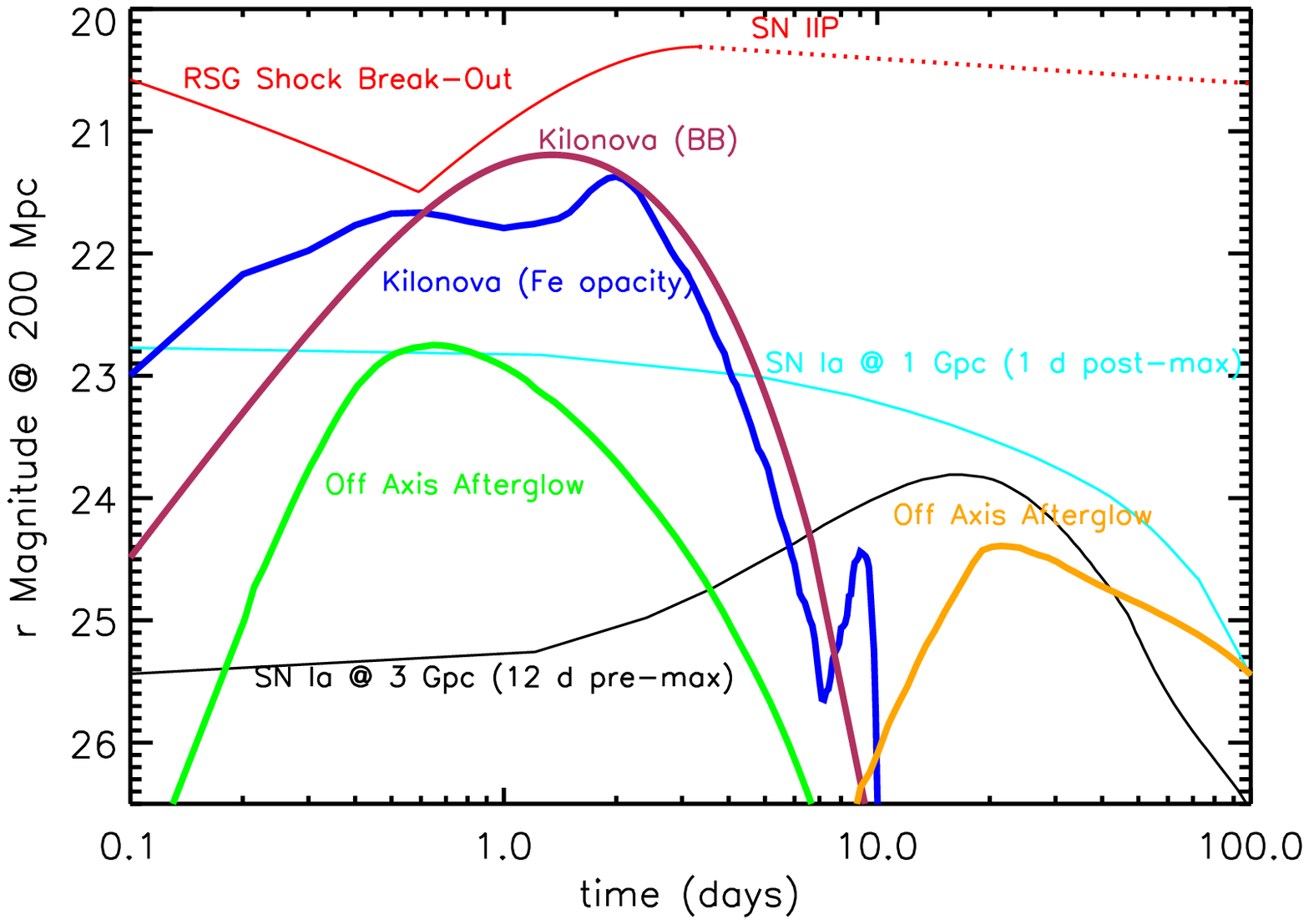} 
\includegraphics[angle=0,width=3.3in]{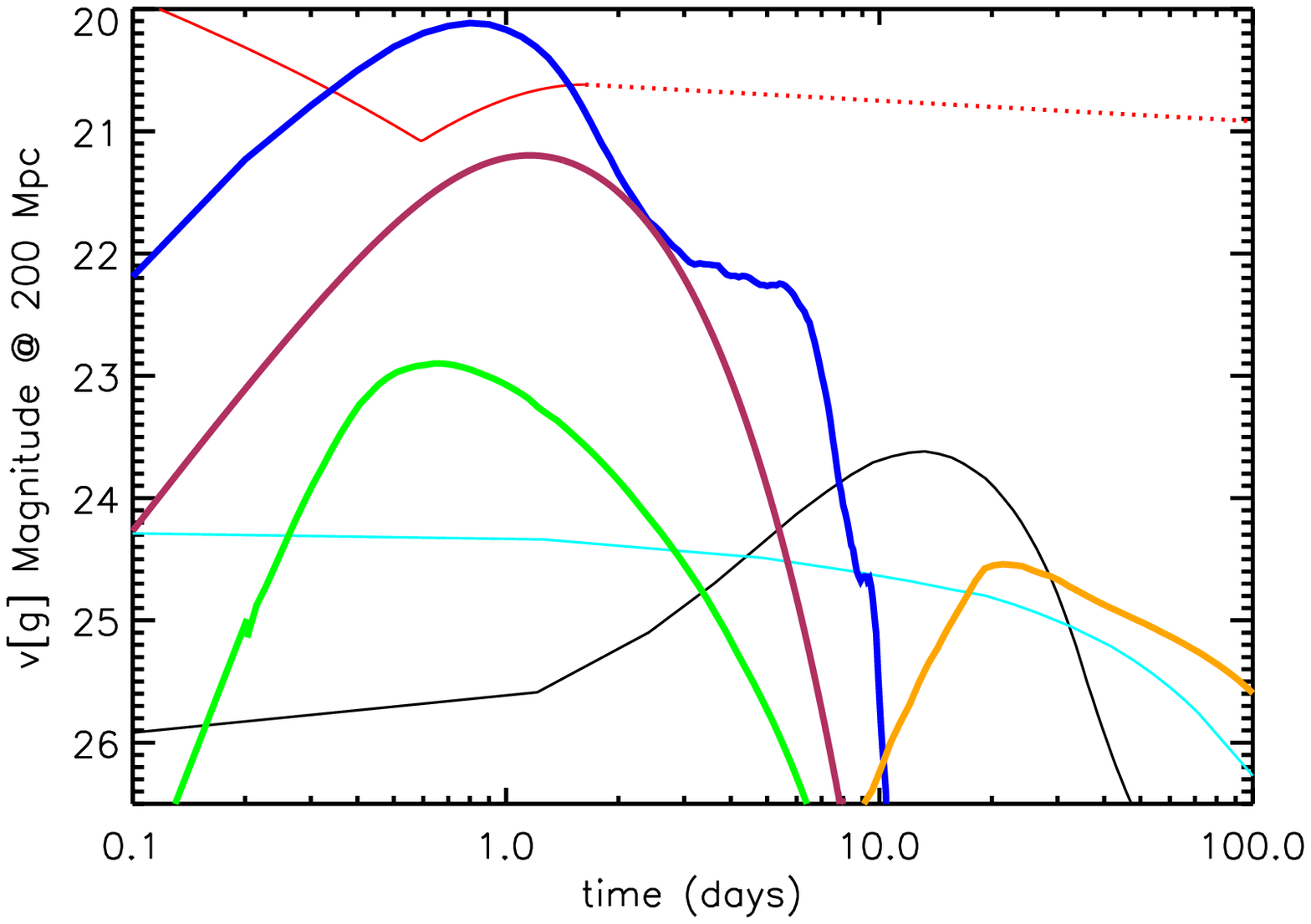}
\includegraphics[angle=0,width=3.3in]{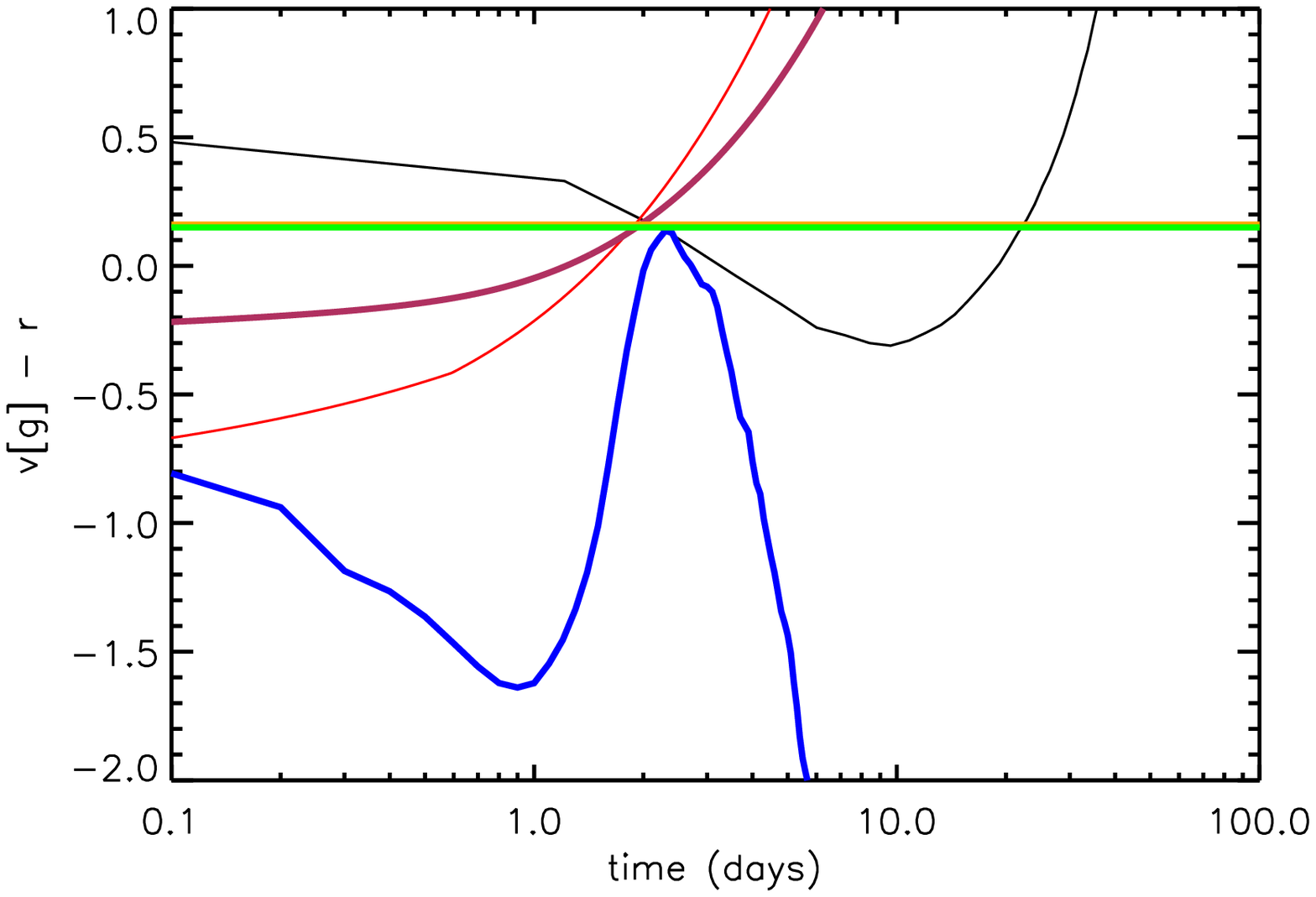}
\includegraphics[angle=0,width=3.3in]{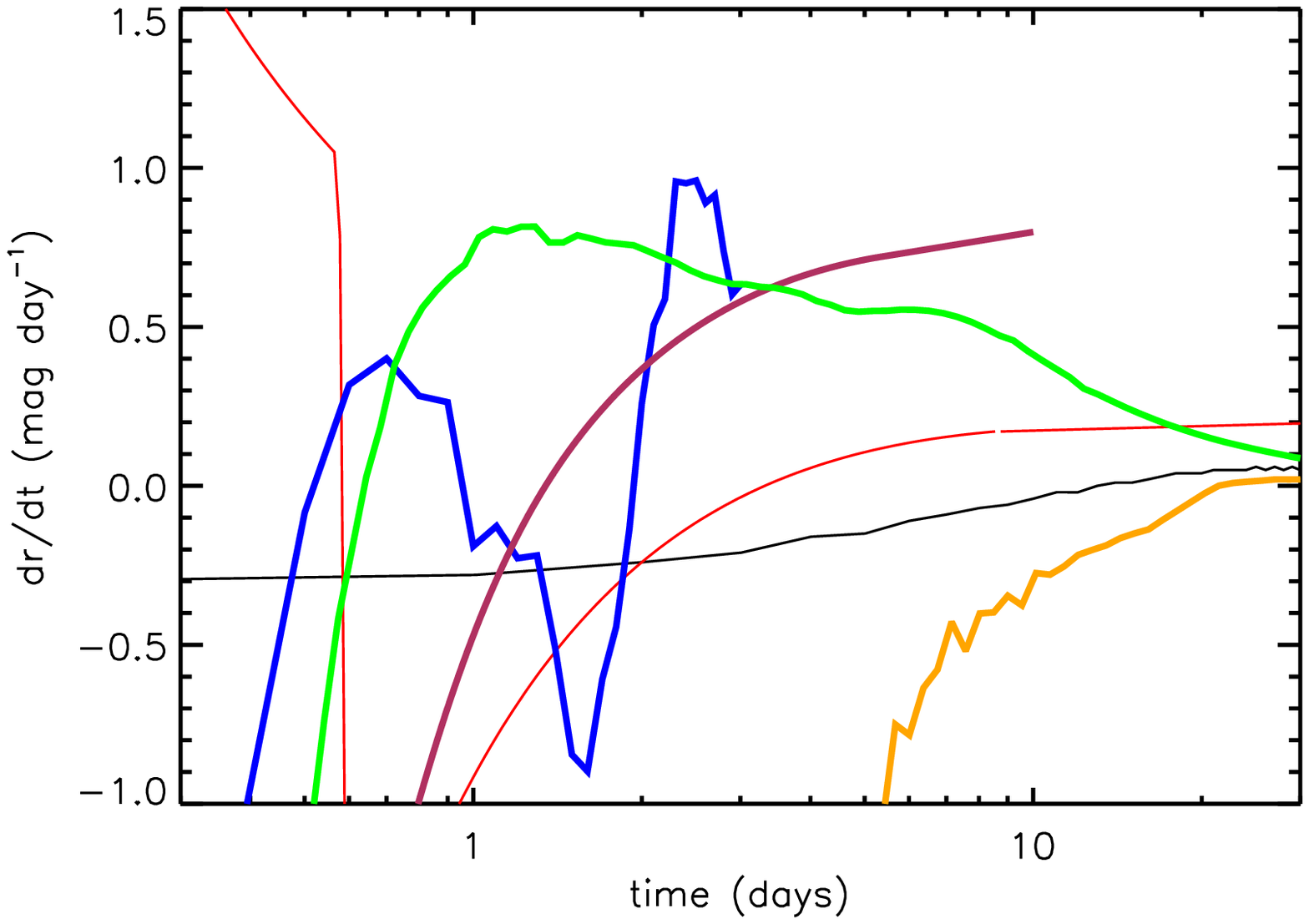} 
\end{center}
\caption[]{Comparison of off-axis afterglow and kilonova light curves
with contaminating optical transients, including $r$-band light curve
(top-left), $V$-band light curve (top-right), $V-r$ color curves
(bottom-left), and $r$-band decline rate (bottom-right).  Off-axis
light curves are shown for the parameters in
Figures~\ref{fig:onaxis}--\ref{fig:offaxis2}.  We use a typical model
for the kilonova light curve, $M_{\rm ej}=10^{-2}$ M$_\odot$ and
$\beta=0.1$, for two different assumptions about the opacity of the
ejecta: the LP98 model assuming black body emission ({\it purple}),
and a model assuming pure Fe opacity from \citet{Metzger+10} ({\it
blue}).  Shown for comparison are background Type Ia supernova light
curves 12 d before maximum ({\it black}) and 1 d after maximum ({\it
cyan}) from \citet{Wang+09}, as well as supernova shock break-out from
a red supergiant ({\it red}) \citep{Nakar&Sari10}, followed by a
standard Type IIP supernova plateau ({\it dashed red}).}
\label{fig:compare}
\end{figure*}

In terms of Virtue \#3, contamination in the radio band is less severe
than in the optical band.  In the optical, we expect contamination
mainly from background supernovae (Type Ia, and Type IIP shock
breakout), while contamination from AGN variability can be avoided
based on its coincidence with a galaxy nucleus.  We evaluate the
off-axis light curves in comparison to supernova light curves in
Figure~\ref{fig:compare}.  For the high density case ($n\sim 1$
cm$^{-3}$), the off-axis light curve peaks on a timescale of $\sim 1$
d, and is indeed similar to the kilonova light curves.  The rapid rise
and decline (declining by $\sim 3$ mag in $\sim 5$ d) can easily
distinguish this case from Type Ia supernovae, which rise and decline
by comparable amounts on timescales of tens of days
(Figure~\ref{fig:compare}).  Shock breakout emission from red
supergiants also leads to rapid rise and decline ($\sim 1$ mag on a
timescale of $\sim 1$ d), but is subsequently followed by a long and
bright plateau phase (a Type IIP supernova) that can easily
distinguish these cases.  Thus, with a sufficiently rapid cadence
(1-day) and a depth similar to the LSST normal survey (or better yet
$\sim 0.5$ hr pointings with $\approx 26.5$ mag), off-axis afterglows
in a dense medium can be separated from background contaminating
supernovae.

The case of an off-axis afterglow in a low density medium ($n\sim
10^{-3}$ cm$^{-3}$) is somewhat more complicated.  The off-axis light
curve peaks on a timescale of $\sim 20$ d, followed by a decline of
about 1 mag in the subsequent 3 months (Figure~\ref{fig:compare}).
With a peak brightness of $\sim 24.5$ mag, a convincing detection
requires a depth beyond the normal LSST survey mode.  However, a 1-day
cadence is not essential, and the depth can be achieved by stacking
multiple images on a $\sim\,$week timescale.  Given the slower
evolution of the light curve, it is more similar to supernova light
curves than the $n\sim 1$ cm$^{-3}$ case.  The predicted rate of
decline is slower than a Type Ia supernova post maximum (i.e.,
$\approx 2.5-3$ mag in $\sim 100$ d).  It is, however, faster than a
typical Type IIP supernova light curve, which exhibits a plateau for
$\sim 100$ d.

Color evolution can in principle also be used to distinguish off-axis
afterglows from other transients.  Since the afterglow is synchrotron
emission (and the optical waveband is generally above the
characteristic frequency, $\nu_m$), it has a power-law spectrum with a
fixed slope $F_{\nu}\propto\nu^{(1-p)/2}$ and hence a constant red
color $g-r\sim 0.2$ mag for $p=2.5$.  By comparison,
Figure~\ref{fig:compare} shows that the color of a shock break-out and
rising IIP SNe increases by a magnitude from blue to red in just a few
days.  Although the colors of a rising SN Ia are similar to the
afterglow emission, events observed near their peak (the case in which
a background Type Ia supernova light curve could be mistaken for a low
density afterglow) are much redder.

Finally, both radio and optical counterparts will satisfy Virtue \#4,
although at low frequency and low signal-to-noise ratio, EVLA/ASKAP
positions will typically be $\gtrsim {\rm few}$ arcsec, as opposed to
sub-arcsecond in the optical band.  At a typical distance of 200 Mpc
this should not be an impediment for a host galaxy association
($1''\approx 0.8$ kpc), but it will not allow a robust study of the
sub-galactic environment, and hence an association with specific
stellar populations (c.f., \citealt{Fong+10}).  It may also impede the
rejection of AGN.

We therefore conclude that optical and radio afterglows do not satisfy
all of the required cardinal Virtues for an EM counterpart.  The
fraction of detectable off-axis optical afterglows is $\sim 0.1$, and
possibly even lower depending on the range of energy and circumburst
density for typical NS-NS/NS-BH mergers.  The fraction of detectable radio
afterglows may be close to zero due to the limited range of $E-n$
phase-space accessible with existing and planned radio telescopes.

\subsection{Kilonova}

If the majority of NS-NS/NS-BH mergers occur in low density
environments ($n \lesssim 10^{-3}$ cm$^{-3}$) or produce low energy
jets ($E\lesssim 10^{49}$ erg), then optical afterglows are no longer
effective counterparts.  This is also true if most NS-NS mergers are
not accompanied by SGRBs.  In these cases, kilonovae provide an
isotropic source of emission that does not depend on the external
environment.  Because the emission is thermal and requires only a
small quantity of neutron-rich ejecta (as is likely to accompany most
mergers), the predicted signal is also relatively robust, and at a
peak optical brightness of $\sim 19-22$ mag is detectable with a
facility such as LSST.  Thus, kilonovae satisfy Virtues \#1, \#2, and
\#4, as long as a rapid and deep search is carried out
(Figure~\ref{fig:LPfrac} and Table~\ref{tab:kilonova}).  Therefore, a
key question is whether these events can be easily distinguished from
contaminating sources.

We present kilonova light curves in comparison to background
supernovae in Figure~\ref{fig:compare}.  We show both a black body
model (LP98) and a model assuming pure Fe opacity \citep{Metzger+10}
to span the plausible range in the true light curve and color
evolution, the latter of which remains especially uncertain due to the
lack of experimental data on the opacity of pure $r$-process ejecta.
As in the case of off-axis afterglow emission in a dense medium, the
kilonova light curve evolution is much more rapid than for supernovae:
the rise time is $\sim 1$ d, followed by a decline of about 3 mag in
$\sim 5-8$ d.  This behavior places stringent constraints on an
effective search (1-day cadence and a depth of $\gtrsim 24$ mag), but
it allows for a clean separation from contaminating supernovae.  Thus,
it appears that kilonovae can satisfy all four Cardinal Virtues.

\subsection{Quality of Information: A Fifth Virtue?}

Even if all types of EM counterparts discussed in this paper will
eventually be detected in conjunction with GW triggers, each provides
distinct information about the merger.  The detection of an SGRB in
temporal coincidence with a GW trigger will establish a firm
connection with NS-NS/NS-BH mergers.  Since these events are also
expected to be face-on mergers, such a detection will establish the
orientation of the binary, thereby allowing for more accurate
extraction of additional binary parameters from the GW signal, such as
the masses and spins of its members and, potentially, information
about the high density equation of state.  While a $\gamma$-ray
detection is not itself likely to substantially improve the positional
accuracy, an on-axis optical afterglow, or a kilonova, should be
detectable and will provide a host galaxy association and redshift.

The detection of an orphan optical afterglow (on- or off-axis) will
provide much of the same information.  Namely, it will establish a
connection with SGRBs and will also possess a nearly face-on
orientation (since orphan optical afterglows are only detectable to
$\lesssim 2\theta_j$).  The brightness of the optical emission will
also provide information on the combination of energy and circumburst
density.  A radio detection will provide no information on binary
orientation (since essentially all off-axis angles are detectable).
The peak time and brightness will provide information on the
combination of energy, density, and ejecta velocity.  As a result of
the velocity degeneracy a radio detection will not necessarily
establish an association with SGRBs (i.e., the specific case of
$\beta\approx 1$).  Finally, kilonova detections provide a unique
probe of the inner workings of the merger, since their light curves
depend on the mass, velocity, and geometry of the ejecta, while their
opacity and spectral features probe the ejecta composition
(spectroscopy will require real-time identification near peak).  The
discovery of a kilonova event will also represent the first in-situ
observation of freshly produced $r$-process material, the origin of
which remains perhaps the biggest mystery in nuclear astrophysics.
However, such detections will not help to establish a connection with
SGRBs.

To summarize, the potential connection of NS-NS mergers with SGRBs
provides a useful EM counterpart for both on- and off-axis emission.
However, in essentially all possible scenarios, only a small fraction
of GW events ($\lesssim 10\%$) will be followed by a detectable SGRB
or optical/radio afterglow.  On the other hand, isotropic kilonovae
will likely provide a larger detection fraction, as long as the
typical ejected mass is $\gtrsim 10^{-3}$ M$_\odot$ and deep
observation with 1-day cadence are carried out.

\section{Follow-Up Strategy Recommendations}
\label{sec:strategy}

Taking advantage of any of the potential EM counterparts discussed in
the previous sections requires a careful observing strategy.  In this
section we make specific recommendations for searches at $\gamma$-ray,
optical, and radio wavelengths to enhance the detection probability,
{\it given a reasonable allocation of resources}.

\subsection{Gamma-Rays}

In the case of $\gamma$-ray detections of an associated SGRB, the
limiting factor is the small fraction of on-axis events within the
ALIGO/Virgo detection volume.  The on-axis orientation provides a
boost to the detection volume, leading to a detectable rate perhaps as
high as $\sim 1$ event per year.  As a result of this low rate,
all-sky coverage in $\gamma$-rays is more critical than the ability to
substantially refine the GW positions.  This will still allow for a
robust association due to the temporal coincidence, and the positional
refinement can then be achieved from the bright on-axis optical
afterglow.  In addition, since a temporal coincidence with an SGRB
allows for a lower threshold GW detection, a strategy of searching the
GW data stream based on $\gamma$-ray triggers may actually lead to a
larger number of detections than the opposite approach (i.e., it will
boost the accessible volume by more than a factor of 3.4 times for
face-on orientation).  Thus, we strongly recommend an operational
$\gamma$-ray satellite in the ALIGO/Virgo era, with capabilities
similar to the {\it Fermi}/GBM.

\subsection{Optical}

The search for orphan optical afterglows requires wide-field
telescopes capable of achieving a depth of at least $\sim 23$ mag, and
perhaps $\sim 26.5$ mag for detections of typical events.  The maximum
achievable depth is determined by the need to cover tens of square
degrees with a 1-day cadence.  Since $\sim 10$ pointings are required
with facilities such as PTF, Pan-STARRS, and LSST, the maximum time
per pointing is about 0.5 hr, leading to depths of $\sim 22$ mag for
PTF, $\sim 23.5$ mag for Pan-STARRS, and $\sim 26.5$ mag for LSST.
For a localization region of a few square degrees (i.e., a single
pointing), the achievable limiting magnitudes are about 1.2 mag
deeper.  The results of Monte Carlo simulations of the detection
fractions in 3 and 5 epochs for on- and off-axis afterglows given a
limiting magnitude and cadence are summarized in
Table~\ref{tab:afterglow} and Figure~\ref{fig:agfrac}.  We find that
only in the case of $E\sim 10^{50}$ erg the standard LSST cadence and
depth are sufficient for multiple detections.  However, for lower
energies (which may be typical of most SGRBs), a faster cadence and
greater depth ($\sim 26.5$ mag) are required for multiple detections.
To achieve a detection fraction of $50\%$ in 3(5) epochs for the case
of $E_j\sim 10^{48}$ erg and $\theta_{\rm obs}=2\theta_j$ requires a
depth of at least $23.5$($26$) mag for a 1-day cadence.  Thus, we
conclude that standard LSST depth and cadence are non-ideal for
detections of off-axis afterglows.

A key issue discussed in previous papers is that at the typical
distance limit of $\sim 200$ Mpc for ALIGO/Virgo detections, one could
expedite the search for EM counterparts by focusing on galaxies within
this volume.  However, the number of galaxies within a typical GW
error region in the typical luminosity range of SGRB hosts ($L\gtrsim
0.1\,L^*$) is\footnotemark\footnotetext{We use the SDSS luminosity
function with $M_r^*\approx -21.2$ mag and $\phi^*\approx 5\times
10^{-3}$ Mpc$^{-3}$ \citep{bhb+03}.  Integration down to $0.1\,L^*$
therefore gives about 8 galaxies per square degree within a distance
of 200 Mpc.} $\sim 400$; the number of $L^*$ galaxies is about 50.
Thus, the number of galaxies is much larger than the required number
of pointings for wide-field telescopes ($\sim 10$), and therefore a
focus on nearby galaxies has no effect on the required cadence and
depth.  Conversely, the number of galaxies is too large for an
efficient search with a large-aperture but small field-of-view
telescope (e.g.,~ Keck, Gemini) since several hundred pointings will
be required within a single night (leading to $\lesssim 1$ min per
pointing).  Even a search of only $L\gtrsim L^*$ galaxies (which will
inevitably miss a substantial fraction of the counterparts in
sub-$L^*$ galaxies) will limit the observations to $\sim 5$ min per
galaxy, and will require the full use of an 8-m class telescope for
several nights.

Thus, our key recommendation is follow-up with wide-field optical
telescopes capable of reaching a depth of $\gtrsim 23$ mag in 0.5 hr,
using a 1-day cadence.  Effectively, this means that LSST should
execute a non-standard cadence to follow up GW triggers (a
``sub-survey'' mode).  Ideally, such dedicated follow-up observations
will also involve longer exposure times than the normal survey mode
(up to $\sim 0.5$ hr per pointing), but even without a change to the
standard exposure time, repeated visits on a nightly basis will
provide the most efficient search strategy for optical counterpart
searches.  The same strategy is key for detections of the
fast-evolving kilonovae (Figure~\ref{fig:LPfrac} and
Table~\ref{tab:kilonova}).  If no convincing counterpart is detected
within a few days, a search for delayed off-axis emission (due to low
density) can employ the normal LSST cadence since the typical
timescale is tens of days (Figure~\ref{fig:agfrac}).

We note that efforts to perform GW-triggered optical follow-up have
already begun during the recent LIGO science run by the LOOC UP
(Locating and Observing Optical Counterparts to Unmodeled Pulses)
project \citep{Kanner+08,Abbott+08}.  LOOC UP reconstructs the sky position of
candidate GW signals to make prompt optical follow-up observations
using wide-field sub-meter class telescopes.  \citet{Cannon+11}
discuss techniques to reduce the latency between GW detection and
follow-up to minutes or even seconds, in which case counterpart
searches could in principle begin simultaneous with the final
coalescence.  The meter class telescopes employed in LOOC UP were
sufficiently sensitive to detect off-axis afterglows or kilonova
within the LIGO volume, but will be clearly insufficient for the
ALIGO/Virgo volume.

\subsection{Radio}

The search for radio counterparts is complicated by the highly
uncertain peak time of the light curve (ranging from months to years),
as well as by the small field-of-view of the EVLA.  For an off-axis
afterglow in a dense medium the expected peak time is $\sim 0.1$ yr,
while for a non-relativistic spherical counterpart the expected peak
occurs on a timescale of a few years.  As a result, a robust search
has to cover a broad range of timescales, with an initial rapid
cadence of $\sim 1$ week, followed by a transition to monthly and then
yearly observations.  In total we estimate that at least $10-15$
epochs will be required, spread logarithmically over a decade.  As
discussed in \S\ref{sec:ra}, a search with the EVLA will require about
30 hr per epoch to cover a typical GW error region.  If we assume that
about half of these epochs will take place within the first year after
the trigger, the required time to follow up the best-bet rate of 40 GW
triggers per year is about 100\% of the EVLA time.

The peak flux density required for a convincing detection (a factor of
2 rise above the achievable $5\sigma=0.25$ mJy threshold) is about
$0.5$ mJy.  As shown in Figure~\ref{fig:fom}, none of the existing
SGRB optical afterglows will lead to such a bright signal.  The
detection of events with $F_{\nu,p}\sim 0.1$ mJy is complicated not
only by the excessive amount of required telescope time (hundreds of
hours per epoch!), but also by source confusion at low frequency, and
the fact that a host galaxy with ${\rm SFR}\sim 1$ M$_\odot$ yr$^{-1}$
has a 1 GHz flux density of $\sim 0.6$ mJy.  These factors will limit
the searches even if the error regions are only a few square degrees.

In terms of the search strategy, at 1 GHz the best approach is to tile
the full error region with $\sim 100-200$ pointings, while at higher
frequencies it is more profitable to target the $\sim 400$ galaxies
with $L\gtrsim 0.1\,L^*$ individually (although this will require even
more observing time).  Observations with a future facility such as
ASKAP will reduce the time requirement to a few hours per epoch by
reducing the number of pointings.  However, the sensitivity of the
search is unlikely to improve since source confusion becomes a
dominant obstacle.

We finally note that for the case of radio emission from
non-relativistic ejecta, the required energy and density are $E\gtrsim
10^{51}$ erg and $n\gtrsim 0.1$ cm$^{-3}$ (Figure~\ref{fig:fom}).  It
is unclear if the required energy scale can be produced in a typical
NS-NS merger, but even if it does, the resulting decade-long delay
between the GW trigger and peak of the putative radio emission will
require many observing epochs, and will furthermore impede a
convincing association.

Thus, our recommendation in the case of radio searches is to limit the
search to a timescale of a few months, appropriate for the case of an
off-axis afterglow ($\beta\sim 1$).  In this case the peak brightness
is also more likely to be detectable, since $F_{\nu,p}\propto
\beta^{11/4}$, and the relatively modest time delay ($t_{\rm dec}
\propto \beta^{-5/3}$) will reduce the potential for contamination.
The required observing time with the EVLA will be about $\sim 200$ hr
for several epochs logarithmically spaced in time over several months.
Future searches with facilities like ASKAP should also focus on the
same timescale since they are limited to the same depth as EVLA.

\section{Conclusions}
\label{sec:conclusion}

With the era of gravitational wave astronomy fast approaching we
investigated and critically assessed a range of potential
electromagnetic counterparts for NS-NS/NS-BH mergers, and their
detectability with existing and upcoming telescopes.  We used the
rates of (on-axis) SGRBs to predict the detection rate with an all-sky
$\gamma$-ray monitor, and existing information about SGRB afterglows
to predict the appearance and brightness of off-axis optical and radio
emission.  Finally, we assessed the light curves and detectability of
kilonovae.  Overall, we found that none of the potential EM
counterparts is guaranteed to satisfy all four Cardinal Virtues ---
detectability, high fraction, identifiability, and positional accuracy
--- but that critical insight into the merger physics can be gained
from any of the counterparts for at least some events.  In particular,
we found that:

\begin{itemize}

\item Gamma-ray observations are critical for establishing a firm
connection between SGRBs and NS-NS/NS-BH mergers.  Such detections are
likely to be limited to a rate of $\lesssim 1$ yr$^{-1}$, but the
face-on configuration will allow for better measurements of the binary
parameters.  In addition, $\gamma$-ray {\it triggered} GW searches may
enhance the probability of joint GW/EM detections.

\item The number of expected galaxies with $L\gtrsim 0.1\,L^*$
(typical of SGRB hosts; \citealt{Berger09}) in a typical GW error
region is $\sim 400$, making targeted searches of galaxies in the
optical and/or radio inefficient.  In both cases, complete coverage of
the error region is less time consuming.

\item On-axis optical emission typical of existing SGRB afterglows is
easily detectable with existing and planned wide-field telescopes at
$\lesssim 200$ Mpc.

\item Off-axis optical afterglow emission is only detectable to
$\theta_{\rm obs}\sim 2\theta_j$, and is hence limited to $\lesssim
10\%$ of all mergers.  Within this range, LSST observations are
required to detect events similar to existing SGRB afterglows, but
with a specialized depth/cadence of $\sim 26.5$ mag (achievable in 0.5
hr) and 1 d.  Observations with the normal LSST survey mode are likely
to miss most counterparts.  With our proposed LSST depth/cadence
contamination from other sources can be avoided based on the rapid
rise and decline time for high density cases, and based on the decline
rate and lack of color evolution for low density cases.

\item Off-axis radio afterglow emission can in principle be detected
at all observer angles, but existing and planned telescopes are
limited to flux levels that exclude the detection of existing SGRB
afterglows.  The long delay relative to the GW trigger (up to many
years) provides an additional obstacle to robust association.  Still,
searches with a weekly-monthly cadence and per-epoch exposure times of
$\sim 30$ hr (EVLA) or $\sim {\rm few}$ hr (ASKAP), may lead to
detections of rare energetic events in a dense medium.  Contamination
from other sources is less severe than in the optical band, but the
positional accuracy of a detection will be poorer than in the optical,
potentially preventing studies of sub-galactic environments.

\item Isotropic optical emission powered by the radioactive decay of
$r$-process elements in the merger ejecta (kilonova) is expected to
reach a peak brightness of $\sim 19-22$ mag at $\Delta t\sim 1$ d,
with a subsequent decline by several magnitudes in a few days.  The
brightness is independent of the ambient density.  The detection of
kilonovae therefore requires rapid cadence ($\sim 1$ d) to at least
the normal LSST survey depth (and preferably a maximal depth of $\sim
26.5$ mag).  The rapid rise and decline will reduce the contamination
from other optical transients (e.g., supernovae).  With the
wherewithal to carry out such a search with LSST, kilonovae can indeed
serve as the most promising counterpart of compact object binary
mergers.

\item Our key recommendations for maximizing the detection probability
of EM counterparts are: An all-sky $\gamma$-ray satellite similar to
{\it Fermi}/GBM; a specialized LSST ``sub-survey'' mode with a 1-day
cadence and a depth of $\sim 26.5$ mag; radio follow-up with a weekly
cadence to a depth of 0.25 mJy using EVLA/ASKAP, limited to $\lesssim
{\rm few}$ months after the trigger.

\end{itemize}

Since the timescales of the various potential EM counterparts are
spread from seconds ($\gamma$-rays) to days (optical) to months
(radio), a staggered approach will clearly inform a joint
observational strategy.  For example, the detection of $\gamma$-ray
emission should trigger an immediate high-cadence search for on-axis
afterglow/kilonova emission in both the optical and radio; such a
search will still require wide-field imaging.  Similarly, the
detection of off-axis afterglow or kilonova optical candidate(s)
should trigger targeted radio observations (allowing for much deeper
observations relative to a complete search of the GW error circle).

We finally note that our framework for evaluating potential EM
counterparts can be revised as the actual positional capabilities of
ALIGO/Virgo and the merger rate become clear.  However, unless the GW
sensitivity or rate have been substantially over-estimated, we
conclude that a concerted follow-up effort will determine whether
NS-NS/NS-BH mergers are associated with SGRBs, and will provide
critical insight into the physics of compact objects and the merger
process.  The fundamental importance of these results justifies the
proposed expensive electromagnetic observational strategy.

\acknowledgements We thank Hendrik van Eerten and Andrew MacFadyen for
producing and maintaining their online library of afterglow light
curves.  We thank Ehud Nakar for providing theoretical light curves of
supernova shock break-out.  We thank S.~Nissanke for helpful discussions and information.  B.D.M.~is supported by NASA through Einstein Postdoctoral Fellowship grant number PF9-00065 awarded by the
Chandra X-ray Center, which is operated by the Smithsonian
Astrophysical Observatory for NASA under contract NAS8-03060.
E.B.~acknowledges support for this work from the National Science
Foundation through Grant AST-1107973.


\clearpage
\begin{deluxetable}{cccccccc}
\tabletypesize{\footnotesize}
\tablecolumns{8}
\tabcolsep0.15in\footnotesize
\tablewidth{0pc}
\tablecaption{Detection Efficiency of Orphan Afterglows within 200 Mpc
\label{tab:afterglow}}
\tablehead {
\colhead{$r_{\rm lim}$} &
\colhead{$\Delta t$}    &
\colhead{$E_{\rm j}$}   &
\colhead{$n$}           &
\colhead{$\theta_{\rm obs}$} &
\colhead{$f_{det,1}$}   &
\colhead{$f_{det,3}$}   &
\colhead{$f_{det,5}$}   \\
\colhead{(AB mag)} &
\colhead{(d)}   &
\colhead{(erg)} &
\colhead{(cm$^{-3}$)} &
\colhead{}    &
\colhead{}    &
\colhead{}    &
\colhead{} 
}
\startdata
22.0 & 1 & $10^{48}$ & 1         & $\theta_j$  & 1.00 & 0.05 & 0.00  \\
23.5 & 1 &\nod       &\nod       &\nod               & 1.00 & 0.40 & 0.02  \\
24.7 & 4 &\nod       &\nod       &\nod               & 1.00 & 0.00 & 0.00  \\
24.7 & 1 &\nod       &\nod       &\nod               & 1.00 & 0.95 & 0.08  \\
26.5 & 1 &\nod       &\nod       &\nod               & 1.00 & 0.95 & 0.80  \\
22.0 & 1 &\nod       &\nod       & $2\theta_j$ & 0.28 & 0.05 & 0.00  \\
23.5 & 1 &\nod       &\nod       &\nod               & 1.00 & 0.37 & 0.07  \\
24.7 & 4 &\nod       &\nod       &\nod               & 0.95 & 0.01 & 0.00  \\
24.7 & 1 &\nod       &\nod       &\nod               & 1.00 & 1.00 & 0.37  \\
26.5 & 1 &\nod       &\nod       &\nod               & 1.00 & 1.00 & 1.00  \\
26.5 & 1 &\nod       &\nod       & $4\theta_j$ & 0.18 & 0.18 & 0.18  \\\hline
22.0 & 1 &\nod       & 10$^{-3}$ & $\theta_j$  & 0.34 & 0.00 & 0.00  \\
23.5 & 1 &\nod       &\nod       &\nod               & 0.56 & 0.01 & 0.00  \\
24.7 & 4 &\nod       &\nod       &\nod               & 0.38 & 0.00 & 0.00  \\
24.7 & 1 &\nod       &\nod       &\nod               & 1.00 & 0.10 & 0.02  \\
26.5 & 1 &\nod       &\nod       &\nod               & 1.00 & 0.93 & 0.29  \\
26.5 & 1 &\nod       &\nod       & 2$\theta_j$ & 0.00 & 0.00 & 0.00  \\\hline
22.0 & 1 & $10^{50}$ & 1         & $\theta_j$  & 1.00 & 1.00 & 1.00  \\
24.7 & 4 &\nod       &\nod       &\nod               & 1.00 & 1.00 & 1.00  \\
22.0 & 1 &\nod       &\nod       & $2\theta_j$ & 1.00 & 1.00 & 1.00  \\
24.7 & 4 &\nod       &\nod       &\nod               & 1.00 & 1.00 & 1.00  \\
22.0 & 1 &\nod       &\nod       & $4\theta_j$ & 0.05 & 0.05 & 0.05  \\
23.5 & 1 &\nod       &\nod       &\nod               & 0.39 & 0.39 & 0.39  \\
24.7 & 4 &\nod       &\nod       &\nod               & 1.00 & 1.00 & 1.00  \\
26.5 & 1 &\nod       &\nod       &\nod               & 1.00 & 1.00 & 1.00  \\\hline
22.0 & 1 &\nod       & $10^{-3}$ & $\theta_j$  & 1.00 & 1.00 & 1.00  \\
24.7 & 4 &\nod       &\nod       &\nod               & 1.00 & 1.00 & 1.00  \\
23.5 & 1 &\nod       &\nod       & $2\theta_j$ & 0.27 & 0.27 & 0.27  \\
24.7 & 4 &\nod       &\nod       &\nod               & 1.00 & 1.00 & 1.00  \\
26.5 & 1 &\nod       &\nod       & $4\theta_j$ & 0.03 & 0.03 & 0.03  
\enddata
\tablecomments{The columns are (left to right): (i) $5\sigma$ $r$-band
limiting magnitude; (ii) observing cadence; (iii) jet energy; (iv)
density; (v) viewing angle; (vi) fraction of events detected in one
epoch; (vii) fraction of events detected in three epochs; and (viii)
fraction of events detected in five epochs.}
\end{deluxetable}

\clearpage
\begin{deluxetable}{cccccccc}
\tabletypesize{\footnotesize}
\tablecolumns{8}
\tabcolsep0.15in\footnotesize
\tablewidth{0pc}
\tablecaption{Detection Efficiency of Kilonovae within 200 Mpc
\label{tab:kilonova}}
\tablehead {
\colhead{$r_{\rm lim}$} &
\colhead{$\Delta t$}    &
\colhead{$M_{\rm ej}$}  &
\colhead{$v_{\rm ej}$}  &
\colhead{$f_{\rm nuc}$}  &
\colhead{$f_{det,1}$}   &
\colhead{$f_{det,3}$}   &
\colhead{$f_{det,5}$}   \\
\colhead{(AB mag)}    &
\colhead{(d)}         &
\colhead{(M$_\odot$)} &
\colhead{($c$)}       &
\colhead{}    &
\colhead{}    &
\colhead{}    &
\colhead{} 
}
\startdata
22.0 & 1 & $10^{-1}$ & 0.1 & $3\times 10^{-6}$   & 1.00 & 1.00 & 1.00 \\ 
24.7 & 4 &\nod       &\nod &\nod                 & 1.00 & 1.00 & 1.00 \\\hline 
22.0 & 1 & $10^{-2}$ &\nod &\nod                 & 1.00 & 0.89 & 0.37 \\ 
23.5 & 1 &\nod       &\nod &\nod                 & 1.00 & 1.00 & 0.96 \\ 
24.7 & 4 &\nod       &\nod &\nod                 & 1.00 & 0.73 & 0.00 \\ 
24.7 & 1 &\nod       &\nod &\nod                 & 1.00 & 1.00 & 1.00 \\ 
23.5 & 1 &\nod       &\nod & $1.5\times 10^{-6}$ & 1.00 & 0.89 & 0.22 \\ 
23.5 & 1 &\nod       &\nod & $6\times 10^{-6}$   & 1.00 & 1.00 & 1.00 \\ 
22.0 & 1 &\nod       & 0.3 & $3\times 10^{-6}$   & 1.00 & 0.89 & 0.03 \\ 
23.5 & 1 &\nod       &\nod &\nod                 & 1.00 & 0.98 & 0.20 \\ 
24.7 & 4 &\nod       &\nod &\nod                 & 0.93 & 0.05 & 0.05 \\ 
24.7 & 1 &\nod       &\nod &\nod                 & 1.00 & 1.00 & 0.73 \\ 
26.5 & 1 &\nod       &\nod &\nod                 & 1.00 & 1.00 & 1.00 \\\hline 
22.0 & 1 & $10^{-3}$ & 0.1 &\nod                 & 0.24 & 0.06 & 0.00 \\ 
23.5 & 1 &\nod       &\nod &\nod                 & 1.00 & 0.47 & 0.00 \\ 
24.7 & 4 &\nod       &\nod &\nod                 & 0.56 & 0.00 & 0.00 \\ 
24.7 & 1 &\nod       &\nod &\nod                 & 1.00 & 0.94 & 0.00 \\ 
26.5 & 1 &\nod       &\nod &\nod                 & 1.00 & 0.98 & 0.00
\enddata
\tablecomments{The columns are (left to right): (i) $5\sigma$ $r$-band
limiting magnitude; (ii) observing cadence; (iii) ejecta mass; (iv)
ejecta velocity; (v) nuclear heating parameter \citep{Li&Paczynski98};
(vi) fraction of events detected in one epoch; (vii) fraction of
events detected in three epochs; and (viii) fraction of events
detected in five epochs.}
\end{deluxetable}

\end{document}